# Preference-Oriented Aggregation of Heterogeneous Distributed Energy Resources for Reserve Dispatch

Jingguan Liu, *Graduate Student Member, IEEE,* Xiaomeng Ai, *Member, IEEE,* Shichang Cui, *Member, IEEE,* Xizhen Xue, *Member, IEEE*, Shengshi Wang, *Member, IEEE*, Jiakun Fang, *Senior Member, IEEE,* Wei Yao, *Senior Member, IEEE*, and Jinyu Wen*, Member, IEEE*

***Abstract*—Aggregating distributed energy resources (DERs) aims to encode their collective flexibility into a single set for efficient grid dispatch. However, existing aggregation methods are overly conservative for heterogeneous DERs due to two main challenges: 1) dimensional heterogeneity, which complicates the combination of flexibilities across different time dimensions, and 2) type heterogeneity, where diverse and irregular DER profiles hinder accurate approximations, resulting in significant flexibility loss. To resolve these challenges, this paper propose a novel preference-oriented aggregation method for reserve dispatch. For dimensional heterogeneity, we extend existing techniques by reformulating the Minkowski sum as a polytope projection problem using a matrix transformation technique. By unifying DERs in a higher-dimensional space and projecting them back into the aggregate feasible region, the proposed technique effectively aggregates dimensionally heterogeneous DERs. For type heterogeneity, we further develop a distributed aggregation-dispatch coordination framework that incorporates reserve dispatch preferences into aggregation. This framework effectively captures the critical, active aggregate flexibility prioritized in optimal reserve dispatch, thereby significantly reducing the flexibility loss when aggregating type-heterogeneous DERs. Numerical tests validate the effectiveness of our method in addressing both heterogeneities and highlight its promising potential for power systems with high reserve requirements.**

## NOMENCLATURE

Main symbols and notations used in this paper are given as follows, with additional ones defined as needed. Scalars are written in italic, matrices and vectors in boldface, and sets in blackboard bold. The set of real numbers and the set of non-negative real numbers are denoted by $\mathbb{R}$ and $\mathbb{R}_+$, respectively. For a matrix $\boldsymbol{A}$, the number of rows is denoted by $\mathrm{row}(\boldsymbol{A})$. For matrices $\boldsymbol{A}$ and $\boldsymbol{B}$, the matrices formed by vertical, horizontal, and block-diagonal stacking are denoted by $[\boldsymbol{A};\boldsymbol{B}]$, $[\boldsymbol{A},\boldsymbol{B}]$, and $\mathrm{blk}(\boldsymbol{A},\boldsymbol{B})$, respectively. The identity matrix and the zero matrix are denoted by $\boldsymbol{I}$ and $\boldsymbol{O}$, respectively.

This work was supported by the National Natural Science Foundation of China under Grant 52177088 and Grant 52207108. *(Corresponding author: Xiaomeng Ai).*

J. Liu, X. Ai, S. Cui, J. Fang, W. Yao and J. Wen are with the State Key Laboratory of Advanced Electromagnetic Technology, Huazhong University of Science and Technology, Wuhan 430074, China (e-mail: jingguanliu@hust.edu.cn; xiaomengai@hust.edu.cn; shichang_cui@hust.edu.cn; jfa@hust.edu.cn; w.yao@hust.edu.cn; jinyu.wen@hust.edu.cn).

S. Wang is with the Engineering Cluster, Singapore Institute of Technology, 828608, Singapore (e-mail: sensewang1997@gmail.com).

X. Xue is with the School of Electrical and Electronic Engineering, Nanyang Technological University, 639798, Singapore (email: xizhen.xue@ntu.edu.sg).

### *A. Abbreviations:*

| | |
|---|---|
| ADMM | Alternating direction method of multipliers. |
| DER | Distributed energy resource. |
| DL | Deferable load. |
| ESS | Energy storage system. |
| EV | Electric vehicle. |
| H-representation | Half-space representation. |
| TCR | Thermal controllable resident. |
| VPP | Virtual power plant. |
| V-representation | Vertex representation. |

### *B. Indices and Sets:*

| | |
|---|---|
| $i/j$ | Bus/Line index. |
| $k/N^K$ | DER index/number. |
| $s/N^S$ | Scenario index/number. |
| $t/N^T$ | Dispatch period index/number. |
| $o$ | Iteration index. |
| $\mathbb{U}^{DER}$ | Individual DER flexibility set. |
| $\mathbb{U}^{agg}$ | Exact aggregate DER flexibility set. |
| $\mathbb{P}^{agg}$ | Approximate aggregate DER flexibility set. |
| $\mathbb{U}^{aff}$ | High-dimensional flexibility set of DERs. |
| $\mathbb{U}^{base}$ | Base set. |

### *C. Parameters:*

| | |
|---|---|
| $\overline{[\cdot]}/\underline{[\cdot]}$ | Maximum/Minimum value of $[\cdot]$. |
| $t^{DER,stt}/t^{DER,end}$ | Available start/end period of DERs. |
| $\bar{\delta}^{DER}$ | Ramping limit of DERs. |
| $\theta^{DER}$ | Dissipation rate of DERs. |
| $\eta^{DER}$ | Conversion coefficient of DERs. |
| $\kappa^{DER,in}/\kappa^{DER,out}$ | Charging/Discharging efficiency of DERs. |
| $w^{DER}$ | Ambient parameter of DERs. |
| $\omega^{DER}$ | Impact factor of the ambient parameter of DERs. |
| $\boldsymbol{H}$ | Left-hand side coefficient matrix of set. |
| $\boldsymbol{h}$ | Right-hand side coefficient vector of set. |
| $\boldsymbol{\Gamma}^{DER}$ | Period transformation matrix of DERs. |
| $\boldsymbol{\Gamma}^{aff}$ | Transformation matrix between $\mathbb{U}^{aff}$ and $\mathbb{U}^{agg}$. |
| $P^{ref}$ | Reference power consumption of DERs when they are not providing reserve to the power system. |
| $C^{gu}/C^{gd}$ | Up/Down reserve capacity price for thermal units. |
| $C^{vu}/C^{vd}$ | Up/Down reserve capacity price for VPPs. |
| $\pi$ | Probability of each scenario. |
| $C^{gud}/C^{gdd}$ | Up/Down reserve deployment price for thermal units. |
| $C^{vud}/C^{vdd}$ | Up/Down reserve deployment price for VPPs. |
| $C^{wc}/C^{ls}$ | Penalty price for wind curtailment/load shedding. |
| $P^{wd,fc}/P^{ld,fc}$ | Forecast scenario of wind power/load demand. |
| $P^{wd,st}/P^{ld,st}$ | Stochastic scenario of wind power/load demand. |
| $S$ | Transformation factor on line. |
| $\bar{P}^{line}$ | Transmission line capacity. |
| $\bar{\delta}^{g}$ | Ramping limit of thermal units. |
| $\boldsymbol{\lambda}$ | Lagrange multiplier associated with the equality constraint on the aggregate set shape parameter during aggregation-dispatch coordination. |

*D. Decision Variables:*

| | |
|---|---|
| $P^{DER,in}/P^{DER,out}$ | Charging/Discharging power of DERs. |
| $E^{DER}$ | Storage energy of DERs. |
| $P^{DER}$ | Power consumption of DERs. |
| $\Gamma^{agg}$ | Scaling factor of VPPs. |
| $\boldsymbol{\gamma}^{agg}$ | Translation vector of VPPs. |
| $P^{agg}$ | Aggregate power consumption of DERs. |
| $\boldsymbol{G}^{aux}/\boldsymbol{\Lambda}^{aux}/\boldsymbol{\beta}^{aux}$ | Auxiliary variable matrix. |
| $C^{PS}$ | Total operation cost of the power system. |
| $C^{f}$ | Fuel cost of thermal units. |
| $P^{gu}/P^{gd}$ | Up/Down reserve capacity of thermal units. |
| $P^{vu}/P^{vd}$ | Up/Down reserve capacity of VPPs. |
| $P^{gud}/P^{gdd}$ | Up/Down reserve deployment of thermal units. |
| $P^{vud}/P^{vdd}$ | Up/Down reserve deployment of VPPs. |
| $P^{wc}/P^{ls}$ | Power of wind curtailment/load shedding. |
| $P^{g,fc}/P^{g,st}$ | Day-ahead/Intra-day generation of thermal units. |

## I. INTRODUCTION

THE large-scale integration of renewables increases volatility and uncertainty in power systems [1], creating a demand for more flexible resources to maintain power balance in grid reserve dispatch [2]. Concurrently, the proliferation of heterogeneous distributed energy resources (DERs)—such as energy storage batteries [3], [4], plug-in electric vehicles (EVs) [5], [6], and heating, ventilation, and air conditioning loads [7], [8]—plays a crucial role in enhancing flexibility and controllability for reserve dispatch. To facilitate the utilization of these DERs, the concept of virtual power plants (VPPs) has emerged, coordinating operations between the power system and local DERs [9], [10]. In this context, VPPs aggregate heterogeneous DERs, representing their individual flexibilities as a single set that captures the collective regulation capabilities, which is then submitted to reserve dispatch [11], [12].

The individual flexibility set of DERs can be represented as a convex subset in half-space representation within the power space [13]. However, the exact calculation of their aggregate flexibility set—the Minkowski sum of the individual flexibility sets—is generally computationally intractable [14]. Therefore, various methods have been developed to efficiently compute approximations of the aggregate flexibility set [15], [16], [17]. Among these, inner approximation methods are notable for ensuring that only feasible points are represented, which is crucial for control applications [18]. Despite several attempts at achieving inner approximations, two challenges still remain:

*1) Challenges of Dimensional Heterogeneity*: Geometric techniques are commonly used to approximate the aggregate flexibility set by creating convex inner approximations of individual flexibility sets using specific convex geometries that facilitate efficient computation of their Minkowski sum, such as box-based [19], [20], zonotope-based [21], ellipsoid-based [22], and polytope-based [13], [18], [23], [24], [25] methods. Among these, polytope-based methods are noted for achieving an excellent balance between accuracy and computational complexity [13]. However, existing polytope-based methods assume that DERs share identical dimensions, limiting their ability to address dimensional heterogeneity—a prevalent issue in real-world DER operations, such as the varying plug-in and plug-out periods of EVs [25]. Despite its common occurrence, this challenge is rarely tackled in the literature. A recent study [25] proposes projecting DERs with varying dimensions onto a common space before applying polytope-based aggregation techniques. This method works when dimensional differences are minor, but it struggles under significant heterogeneity. For example, in a 24-hour scheduling scenario, if one DER is active from 1:00 to 7:00 and another from 6:00 to 14:00, the overlapping period is only 6:00–7:00, yielding a mere two-dimensional aggregated region. In contrast, the true aggregate feasible region should extend from 1:00 to 14:00, encompassing 14 dimensions, which illustrates the overly conservative nature of the method. Moreover, if the second DER's active period shifts to 8:00–14:00, no common dimension exists, resulting in a collapsed (empty) feasible region. Thus, effectively aggregating dimensionally heterogeneous DERs remains an unresolved challenge.

*2) Challenges of Type Heterogeneity*: Existing inner aggregation methods commonly maximize the volume of aggregate set and have demonstrated satisfactory aggregation accuracy for the same type of DERs [8]. However, when applied to type-heterogeneous DERs, these methods tend to lose significant flexibility due to diverse and irregular DER profiles, which may lead to overly conservative aggregation results [13]. This raises an important question: is perfect approximation accuracy always necessary? We briefly answer this question from two perspectives: *(i) Feasibility*: Type heterogeneity results in a wide range of DER profiles, making it difficult to approximate different types with a single uniform prototype [26]. As a result, some flexibility loss during approximation is inevitable [27]. *(ii) Necessity*: In practice, the aggregate flexibility set submitted for reserve dispatch usually covers only a subset of the regions that are actively used in the optimal grid reserve dispatch [28]. This suggests that the exact aggregate set often includes many inactive areas that are irrelevant for the optimal reserve dispatch. Recognizing this, if we shift our focus from merely maximizing aggregate volume to capturing the critical, active regions preferred by optimal reserve dispatch, we can achieve a high-quality aggregate result that significantly reduces the conservatism introduced by type heterogeneity. Despite its potential, this promising approach remains largely unexplored in the literature.

In light of the above research gaps, this paper pioneers the exploration of a preference-oriented aggregation method for heterogeneous DERs. The main contributions are as follows:

1) ***New technique***: In contrast to existing polytope-based aggregation methods [13], [18], [23], [24], [25], we generalize them by reformulating the Minkowski sum as a polytope projection problem via a matrix transformation technique. This technique unifies DERs in a higher-dimensional space before projecting them back into the aggregate feasible region, thereby overcoming the challenges of direct summation and effectively aggregating dimensionally heterogeneous DERs—a critical step for practical implementation. Also, we present a linear reformulation to ensure computational tractability.

2) ***New insight***: To the best of our knowledge, this is the first time to propose a preference-oriented insight for addressing type heterogeneity in DER aggregation. By employing a distributed aggregation-dispatch coordination framework, our method captures the critical, active aggregate flexibility prioritized in optimal reserve dispatch. Therefore, it produces high-quality aggregate results and significantly mitigates the impact of flexibility loss that often seen in conventional preference-agnostic strategies for type heterogeneous DERs. Furthermore, we develop a ADMM algorithm with parallel-enabled property and warm-start acceleration strategy to efficiently solve the framework.

Following is the remainder of this paper. Section II outlines the problem formulation. Section III describes the generalized inner approximation technique. Section IV discusses the preference-oriented aggregation framework. Section V presents case studies with conclusion in Section VI.

## II. Problem Formulation

In this section, we introduce generalized operation models for various types of DERs and clarify that the main focus of this paper is on finding inner approximations of the aggregate flexibility set of these DERs.

### A. Generalized DER Model

The power system integrates various DER types, including energy storage systems (ESSs) like batteries [29], thermal controllable residents (TCRs) such as air conditioners [7], and deferrable loads (DLs) like plug-in EVs [25]. This paper focuses on these three types, with detailed models provided in Appendix A. The dispatch horizon is discrete, comprising $N^T$ periods of equal length $\Delta t$. Drawing from [13], the operational model for a single DER with $N_{i,k}^{AP}$ active periods is uniformly using the following components: power limits (1.a)-(1.b), storage energy limits (1.c), ramping limits (1.d), the relationship between storage energy and power input/output (1.e), and the power consumption relationship (1.f). Other DER types with simpler models, such as distributed gas turbines, can also be represented using (1) by partially omitting the temporal constraints (1.d)-(1.e). Due to space constraints, we elaborate further on these three types as examples.

$$\underline{P}_{i,k}^{DER,in} \le P_{i,k,t}^{DER,in} \le \overline{P}_{i,k}^{DER,in}, \forall t_{i,k}^{DER,stt} \le t \le t_{i,k}^{DER,end} \tag{1.a}$$

$$\underline{P}_{i,k}^{DER,out} \le P_{i,k,t}^{DER,out} \le \overline{P}_{i,k}^{DER,out}, \forall t_{i,k}^{DER,stt} \le t \le t_{i,k}^{DER,end} \tag{1.b}$$

$$\underline{E}_{i,k}^{DER} \le E_{i,k,t}^{DER} \le \overline{E}_{i,k}^{DER}, \forall t_{i,k}^{DER,stt} \le t \le t_{i,k}^{DER,end} \tag{1.c}$$

$$-\overline{\delta}_{i,k}^{DER} \le \left\{ P_{i,k,t-1}^{DER,in} - P_{i,k,t}^{DER,in}, P_{i,k,t-1}^{DER,out} - P_{i,k,t}^{DER,out} \right\} \le \overline{\delta}_{i,k}^{DER}, \forall t_{i,k}^{DER,stt} \le t \le t_{i,k}^{DER,end} \tag{1.d}$$

$$E_{i,k,t}^{DER} = \theta_{i,k}^{DER} E_{i,k,t-1}^{DER} + \eta_{i,k}^{DER} \left( \kappa_{i,k}^{DER,in} P_{i,k,t}^{DER,in} - P_{i,k,t}^{DER,out} / \kappa_{i,k}^{DER,out} \right) \Delta t + \omega_{i,k}^{DER} w_{i,k}^{DER}, \forall t_{i,k}^{DER,stt} \le t \le t_{i,k}^{DER,end} \tag{1.e}$$

$$P_{i,k,t}^{DER} = P_{i,k,t}^{DER,in} - P_{i,k,t}^{DER,out} \tag{1.f}$$

By further treating net power consumption $P_{i,k,t}^{DER}$ as controllable variables, (1.a)-(1.f) can be simplified into the half-space representation (H-representation) of a full-dimensional polytope $\mathbb{U}_{i,k}^{DER}$:

$$\mathbb{U}_{i,k}^{DER} := \left\{ \boldsymbol{P}_{i,k}^{DER} \in \mathbb{R}^{N_{i,k}^{AP}} \mid \boldsymbol{H}_{i,k}^{DER} \boldsymbol{P}_{i,k}^{DER} \le \boldsymbol{h}_{i,k}^{DER} \right\} \tag{2}$$

where $\boldsymbol{H}_{i,k}^{DER}/\boldsymbol{h}_{i,k}^{DER}$ is the known coefficient matrix/vector, with detailed derivation provided in Appendix B; the bold variable vector $\boldsymbol{P}_{i,k}^{DER}$ represents the collection of italic variables $P_{i,k,t}^{DER}$ arranged in time sequence, and subsequent bold variable vectors represent similar collections of their italic counterparts.

### B. Aggregate Flexibility Set

The aggregate flexibility set for $N_i^K$ DERs within the VPP $i$ can be expressed as the Minkowski sum of their individual flexibility sets [13], denoted by $\mathbb{U}_i^{agg}$ as below:

$$\mathbb{U}_i^{agg} := \biguplus_k \mathbb{U}_{i,k}^{DER} = \left\{ \boldsymbol{P}_i^{agg} \in \mathbb{R}^{N^T} \mid \boldsymbol{P}_i^{agg} = \sum_k \boldsymbol{\Gamma}_{i,k}^{DER} \boldsymbol{P}_{i,k}^{DER}, \boldsymbol{P}_{i,k}^{DER} \in \mathbb{U}_{i,k}^{DER} \right\} \tag{3}$$

where $\uplus$ represents the Minkowski sum computation; $\boldsymbol{P}_i^{agg}$ is the variable vector of aggregated power consumption; $\boldsymbol{\Gamma}_{i,k}^{DER} \in \mathbb{R}^{N^T \times N_{i,k}^{AP}}$ is the period transformation matrix representing the coupling relationships between time periods as shown in (4), where $\boldsymbol{O}_{i,k}^{DER,lhs} \in \mathbb{R}^{(t_{i,k}^{DER,stt}-1) \times N_{i,k}^{AP}}$ is the left-hand side zero matrix, $\boldsymbol{I}_{i,k}^{DER} \in \mathbb{R}^{N_{i,k}^{AP} \times N_{i,k}^{AP}}$ is identity matrix, and $\boldsymbol{O}_{i,k}^{DER,rhs} \in \mathbb{R}^{(N^T - t_{i,k}^{DER,end}) \times N_{i,k}^{AP}}$ is the right-hand side zero matrix.

$$\boldsymbol{\Gamma}_{i,k}^{DER} = \left[ \boldsymbol{O}_{i,k}^{DER,lhs}; \boldsymbol{I}_{i,k}^{DER}; \boldsymbol{O}_{i,k}^{DER,rhs} \right] \tag{4}$$

However, it should be noted that computing the Minkowski sum of two H-representation polytopes is generally NP-hard [24]. Although obtaining the Minkowski sum of convex polytopes in vertex representation (V-representation) is straightforward, converting between H-representation and V-representation (i.e., vertex and facet enumeration) incurs worst-case complexities that grow exponentially with the number of dimensions [30]. Since DER flexibility sets are typically high-dimensional H-representation polytopes as shown in (2), deriving their aggregated flexibility exactly becomes computationally intractable—especially for DER polytopes with diverse parameters (i.e., types) and variables across different time dimensions. Consequently, the NP-hard nature of computing the exact Minkowski sum for arbitrary facet-defined polytopes renders it impractical for real-world applications. For a more detailed theoretical analysis of the Minkowski sum, interested readers may refer to [30], [31].

Recognizing this challenge, the primary goal of this paper is to develop an efficient method to identify an inner approximate set $\mathbb{P}_i^{agg}$ such that:

$$\mathbb{P}_i^{agg} \subseteq \mathbb{U}_i^{agg} \tag{5}$$

With this task in mind, we will address two major challenges of dimensional and type heterogeneity in the following Section III and IV, respectively.

**Remark 1** (Disaggregation Feasibility). Following the results of grid reserve dispatch, the VPPs receive the scheduled value of aggregate power $\boldsymbol{P}_i^{agg,*} \in \mathbb{P}_i^{agg} \subseteq \mathbb{U}_i^{agg}$, which is then disaggregated among the heterogeneous DERs. The disaggregation process for each realization of DERs can be expressed as:

$$\begin{aligned} &\min\ \mathcal{C}(\boldsymbol{P}_{i,k}^{DER,in}, \boldsymbol{P}_{i,k}^{DER,out}, \boldsymbol{P}_{i,k}^{DER}, \boldsymbol{E}_{i,k}^{DER}) \\ &\text{s.t.}\ \ \boldsymbol{P}_i^{agg,*} = \sum_k \boldsymbol{P}_{i,k}^{DER}, \boldsymbol{P}_{i,k}^{DER} \in \mathbb{U}_{i,k}^{DER} \end{aligned} \tag{6}$$

where $\mathcal{C}(\cdot)$ denotes the DER operational cost, represented as an abstract function. Depending on the specific operational requirements, this cost function can be defined as quadratic, linear, or take another appropriate form [32], [33]. By solving (6), we obtain a feasible disaggregation strategy for DERs that strictly complies with their operational constraints, as the proposed inner approximation guarantees disaggregation feasibility for any profile within the aggregate flexibility set $\mathbb{P}_i^{agg}$. However, our primary focus is on deriving the aggregate flexibility set for DERs; suitable disaggregation strategies are left for future research.

## III. Generalized Inner Approximation Technique

In this section, we generalize the existing polytope-based inner approximation methods by reformulating the Minkowski sum as a polytope projection problem using a matrix transformation technique. This technique addresses the dimensional collapse induced by dimensional heterogeneity. We further show that this inner approximation can be recast as a set of linear constraints, enabling tractable computation.

### A. Main Result

Traditional polytope-based inner approximation techniques work at the DER level by first approximating each DER individually and then computing their Minkowski sum in closed-form. However, due to the varying dimensions of DERs, no single prototype can consistently approximate them. This limitation, further discussed in **Remark 2**, motivates our proposed method.

In contrast, our technique adopts a global perspective by aggregating DERs at the VPP level rather than treating each DER individually. We first construct a high-dimensional space spanned by the heterogeneous DERs, then project the resulting polytope onto the lower-dimensional aggregate set of the VPP. This approach offers two key advantages. First, it enables a uniform description and management of dimensionally heterogeneous DERs in the high-dimensional space, thereby overcoming the limitations of individual approximations. Second, the projection can be efficiently formulated using a matrix transformation, thereby sidestepping the complications of cumulatively aggregating inconsistent approximations. This core idea underpins our proposed technique.

The overall aggregation process of our technique is illustrated in Fig. 1, and the detailed steps are provided below.

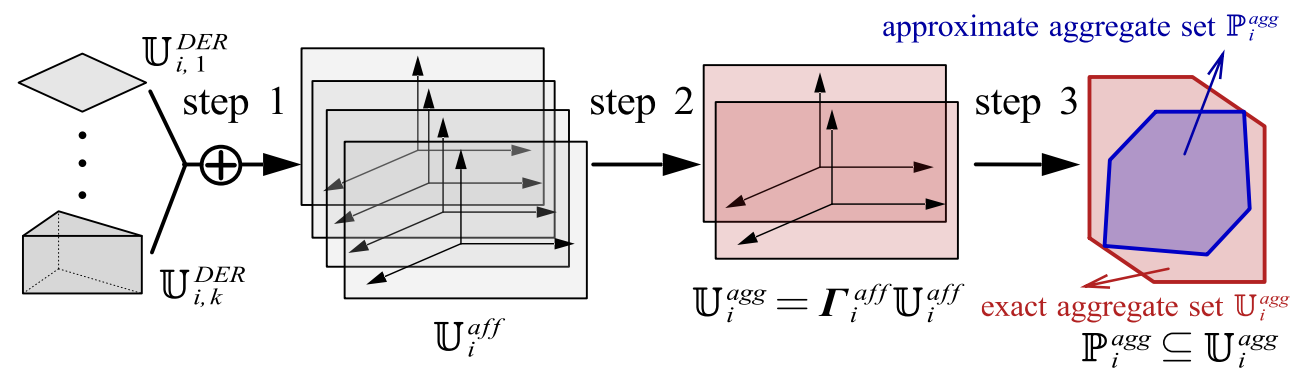

Fig. 1. Illustration of aggregation process.

Three key steps are proposed to efficiently derive $\mathbb{P}_i^{agg}$:

**Step 1** (lifting to a high-dimensional space): First, we construct a high-dimensional polytope $\mathbb{U}_i^{aff}$ to encapsulate the dimensionally heterogeneous DERs within a single VPP, as shown in (7).

$$\mathbb{U}_i^{aff} := \left\{ \boldsymbol{P}_i^{aff} \in \mathbb{R}^{N_i^{AP,total}} \mid \boldsymbol{H}_i^{aff}\boldsymbol{P}_i^{aff} \le \boldsymbol{h}_i^{aff} \right\} \tag{7}$$

where $N_i^{AP,total} = N_{i,1}^{AP} + N_{i,2}^{AP} + \ldots + N_{i,N_i^K}^{AP}$, $\boldsymbol{H}_i^{aff} = \text{blk}\left(\boldsymbol{H}_{i,1}^{DER}, \boldsymbol{H}_{i,2}^{DER}, \ldots, \boldsymbol{H}_{i,N_i^K}^{DER}\right)$, $\boldsymbol{h}_i^{aff} = \left[\boldsymbol{h}_{i,1}^{DER}; \boldsymbol{h}_{i,2}^{DER}; \ldots; \boldsymbol{h}_{i,N_i^K}^{DER}\right]$, and $\boldsymbol{P}_i^{aff} = \left[\boldsymbol{P}_{i,1}^{DER}; \boldsymbol{P}_{i,2}^{DER}; \ldots; \boldsymbol{P}_{i,N_i^K}^{DER}\right]$.

**Step 2** (projection into the aggregate power space): Combining (3) and (7) yields the relationship between $\boldsymbol{P}_i^{agg}$ and $\boldsymbol{P}_i^{aff}$ as follows:

$$\boldsymbol{P}_i^{agg} = \sum_k \boldsymbol{\Gamma}_{i,k}^{DER}\boldsymbol{P}_{i,k}^{DER} = \boldsymbol{\Gamma}_i^{aff}\boldsymbol{P}_i^{aff} \tag{8}$$

where $\boldsymbol{\Gamma}_i^{aff} := \left[\boldsymbol{\Gamma}_{i,1}^{DER}, \boldsymbol{\Gamma}_{i,2}^{DER}, \ldots, \boldsymbol{\Gamma}_{i,N_i^K}^{DER}\right] \in \mathbb{R}^{N^T \times N_i^{AP,total}}$ is the projection matrix.

We now reinterpret equality (8) through the lens of polytope projection [17]. Specifically, the exact aggregate set $\mathbb{U}_i^{agg}$ is reformulated as the projection of the high-dimensional polytope $\mathbb{U}_i^{aff}$, as shown in (9). In contrast to formulation (3) at the DER level, which aggregates DERs by direct summation across dimensions, the VPP-level formulation in (9) naturally avoids this issue by employing the projection matrix $\boldsymbol{\Gamma}_i^{aff}$.

$$\mathbb{U}_i^{agg} = \boldsymbol{\Gamma}_i^{aff}\mathbb{U}_i^{aff} \tag{9}$$

**Step 3** (inner approximation for explicit representation): Combining (5) and (9), the inner approximation process can be reformulated as an inclusion problem between polytope $\mathbb{P}_i^{agg}$ and the projection of polytope $\mathbb{U}_i^{aff}$ by matrix $\boldsymbol{\Gamma}_i^{aff}$.

$$\mathbb{P}_i^{agg} \subseteq \mathbb{U}_i^{agg} = \boldsymbol{\Gamma}_i^{aff}\mathbb{U}_i^{aff} \tag{10}$$

To enhance the efficiency of deriving an explicit analytical representation of the polytope $\mathbb{P}_i^{agg}$, we draw inspiration from existing polytope-based methods [25]. Specifically, we focus on approximating $\mathbb{P}_i^{agg}$ that result from the scaling and translation of a specified full-dimensional set $\mathbb{U}_i^{base}$, namely:

$$\mathbb{P}_i^{agg} = \boldsymbol{\gamma}_i^{agg} + \Gamma_i^{agg}\mathbb{U}_i^{base}, \mathbb{U}_i^{base} = \left\{\boldsymbol{P}_i^{base} \in \mathbb{R}^{N^T} \mid \boldsymbol{H}_i^{base}\boldsymbol{P}_i^{base} \le \boldsymbol{h}_i^{base}\right\} \tag{11}$$

where $\Gamma_i^{agg} \in \mathbb{R}^1$ and $\boldsymbol{\gamma}_i^{agg} \in \mathbb{R}^{N^T}$ are the scaling factor and translation vector, respectively, both of which require proper selection to form $\mathbb{P}_i^{agg}$; The set $\mathbb{U}_i^{base}$ is referred to as the base set. Its coefficient matrix $\boldsymbol{H}_i^{base}$ and vector $\boldsymbol{h}_i^{base}$ are predefined known parameters. For DERs of the same type, the base set parameters can be determined by averaging the parameters of all internal DERs within the VPP [13].

Given the known parameters of DERs, including $\boldsymbol{H}_i^{aff}$, $\boldsymbol{h}_i^{aff}$, and $\boldsymbol{\Gamma}_i^{aff}$, as well as the known parameters of the base set, $\boldsymbol{H}_i^{base}$ and $\boldsymbol{h}_i^{base}$, we can treat $\Gamma_i^{agg}$ and $\boldsymbol{\gamma}_i^{agg}$ as decision variables to reformulate (10) as the following containment constraints:

$$\text{s.t. } \boldsymbol{\gamma}_i^{agg} + \Gamma_i^{agg}\mathbb{U}_i^{base} \subseteq \boldsymbol{\Gamma}_i^{aff}\mathbb{U}_i^{aff} \tag{12}$$

**Remark 2** (Comparison to existing polytope-based methods). The existing polytope-based methods [13], [18], [23], [24], [25] construct an inner approximation of each DER's feasible set by solving (13):

$$\max_{\Gamma_{i,k}^{agg},\boldsymbol{\gamma}_{i,k}^{agg}} \Gamma_{i,k}^{agg} \quad \text{s.t. } \boldsymbol{\gamma}_{i,k}^{agg} + \Gamma_{i,k}^{agg}\mathbb{U}_i^{base} \subseteq \mathbb{U}_{i,k}^{DER} \tag{13}$$

These individual approximations take the form $\boldsymbol{\gamma}_{i,k}^{agg} + \Gamma_{i,k}^{agg}\mathbb{U}_i^{base}$, where $\mathbb{U}_i^{base}$ is a common base set. The aggregate flexibility set is then approximated by summing these affine-transformed sets, expressed as $\biguplus_k\left(\boldsymbol{\gamma}_{i,k}^{agg} + \Gamma_{i,k}^{agg}\mathbb{U}_i^{base}\right) = \left(\sum_k \boldsymbol{\gamma}_{i,k}^{agg}\right) + \left(\sum_k \Gamma_{i,k}^{agg}\right)\mathbb{U}_i^{base} \subseteq \mathbb{U}_i^{agg}$. It can be seen that this approach effectively converts the NP-hard Minkowski sum into a closed-form and computationally efficient algebraic expression. However, its accuracy is limited when the DERs involved are dimensionally heterogeneous.

The conservatism arises from the challenge of finding a universal base set $\mathbb{U}_i^{base}$ that adequately captures the diversity in period dimensions across DER flexibility sets. In cases where a DER set has a lower dimension than the base (i.e., $\dim\mathbb{U}_{i,k}^{DER} < \dim\mathbb{U}_i^{base}$), the only feasible solution to (13) results in a zero scaling factor ($\Gamma_{i,k}^{agg} = 0$), causing dimensional collapse. On the other hand, when a DER set has a higher dimension than the base set (i.e., $\dim\mathbb{U}_{i,k}^{DER} > \dim\mathbb{U}_i^{base}$), the approximation becomes a lower-dimensional subset of the original set, substantially underestimating available flexibility—such as reducing a three-dimensional feasible region to one dimension. Although recent work [25] attempts to mitigate this issue by projecting all DER sets onto a common subspace for polytope-based methods, such a reduction inevitably results in flexibility loss. Moreover, if no shared subspace exists among the DERs, dimensional collapse remains unavoidable, as illustrated by the example presented earlier in the Introduction part.

In summary, for DERs with varying dimensions, existing polytope-based methods are inherently constrained by the need to approximate each dimensionally heterogeneous set using a single base set, resulting in overly conservative outcomes due to dimensional incompatibility.

In contrast, the proposed matrix transformation-based approach addresses this limitation by reframing the aggregation process. Instead of forcing all DERs into a shared dimensional subspace, we interpret the Minkowski sum as a two-step procedure: lifting the flexibility sets into a higher-dimensional space, followed by a structured reduction. Crucially, this method only requires that the union of DER dimensions be compatible with the base set, rather than their intersection—a much less restrictive condition that inherently avoids dimensional collapse. This strategy enables a more accurate and less conservative inner approximation of the aggregate flexibility set, especially in DERs with high dimensional heterogeneity. Moreover, the proposed method encompasses existing polytope-based approaches as a special case: when $\boldsymbol{\Gamma}_i^{aff} = \boldsymbol{\Gamma}_{i,N_i^K}^{DER}, N_i^K = 1$, our formulation reduces precisely to the traditional approach.

Overall, the above comparative analysis underscores the proposed method's unique effectiveness in addressing dimensional heterogeneity compared to existing approaches.

### B. Tractable Reformulation

However, the above containment constraint (12) is difficult to express in an analytical form suitable for optimization, and therefore cannot be directly handled by standard commercial solvers. Fortunately, inspired by Farkas' Lemma for convex polytopes [30], [34], [35], **Proposition 3** is proposed to recast (12) into a solvable linear from for tractable computation, as shown in (14.a)-(14.b).

**Proposition 3** (Containment constraint reformulation). It holds that $\boldsymbol{\gamma}_i^{agg} + \Gamma_i^{agg}\mathbb{U}_i^{base} \subseteq \boldsymbol{\Gamma}_i^{aff}\mathbb{U}_i^{aff}$ if there exist identity matrix $\boldsymbol{I}_i^{agg} \in \mathbb{R}^{N^T\times N^T}$ and auxiliary variables $\boldsymbol{G}_i^{aux} \in \mathbb{R}^{N_i^{AP,total}\times N^T}$, $\boldsymbol{\Lambda}_i^{aux} \in \mathbb{R}_+^{\mathrm{row}\left(\boldsymbol{H}_i^{aff}\right)\times\mathrm{row}\left(\boldsymbol{H}_i^{base}\right)}$, and $\boldsymbol{\beta}_i^{aux} \in \mathbb{R}^{N_i^{AP,total}}$ such that the following relations hold:

$$\Gamma_i^{agg}\boldsymbol{I}_i^{agg} = \boldsymbol{\Gamma}_i^{aff}\boldsymbol{G}_i^{aux}, -\boldsymbol{\gamma}_i^{agg} = \boldsymbol{\Gamma}_i^{aff}\boldsymbol{\beta}_i^{aux} \tag{14.a}$$

$$\boldsymbol{\Lambda}_i^{aux}\boldsymbol{H}_i^{base} = \boldsymbol{H}_i^{aff}\boldsymbol{G}_i^{aux}, \boldsymbol{\Lambda}_i^{aux}\boldsymbol{h}_i^{base} \le \boldsymbol{h}_i^{aff} + \boldsymbol{H}_i^{aff}\boldsymbol{\beta}_i^{aux} \tag{14.b}$$

*Proof*. See Appendix C.

It can be seen that constraints (14.a)-(14.b) provide a tractable linear representation of the original complex containment constraint (12). Although numerical simulations later demonstrate that the proposed linear reformulation achieves satisfactory computational efficiency for aggregating heterogeneous DERs, the reformulation's complexity is not yet perfect. In particular, the number of continuous variables and constraints in (14.a)-(14.b) scales with the number of DERs, causing the overall problem size to grow with the input DER data, raising concerns for large-scale implementation.

Nevertheless, we note that the proposed method naturally supports large-scale aggregation. While its formulation as a linear programming problem enables commercial solvers like GUROBI [36] to handle sizable instances, the greater advantage lies in our extension of existing polytope-based methods. By leveraging the established grouping strategies in existing polytope-based methods, the proposed method can further reduce computational complexity. For example, 1,000 DERs can be partitioned into five groups of 200, with each group's flexibility aggregated in parallel before combining the results. Notably, if each DER were treated as an individual group, our method would revert to existing polytope-based methods and encounter dimensional collapse when aggregating heterogeneous DERs, as noted in **Remark 2**. Thus, our extension not only overcomes the dimensional challenges associated with individual DER aggregation but also enables a more flexible grouping strategy to harness the benefits of grouping and achieve scalability—something that conventional methods, which assign each DER to its own group, cannot offer. Overall, the above analysis highlights the computational efficiency of the proposed linear reformulation relative to traditional methods.

With the above linear reformulation in hand, we then focus on preference-oriented aggregation in the next section to guide the optimal selection of $\boldsymbol{\gamma}_i^{agg}$ and $\Gamma_i^{agg}$.

## IV. Preference-Oriented Aggregation Method

Existing inner approximation methods typically maximize the volume of the approximate aggregate set to determine the optimal values of $\boldsymbol{\gamma}_i^{agg}$ and $\Gamma_i^{agg}$ (as noted in **Remark 2**). However, when applied to type-heterogeneous DERs, these methods often sacrifice much flexibility due to the diverse and irregular shapes of individual DER sets, resulting in overly conservative aggregation as highlighted earlier in the Introduction part. While some flexibility loss during approximation is inevitable, this section develops a preference-oriented aggregation method that focuses on capturing the critical, active aggregate feasible regions preferred by optimal reserve dispatch. This approach yields high-quality aggregate results and significantly reduces conservatism often seen in conventional preference-agnostic methods.

To achieve this, we will incorporate reserve dispatch preferences into the optimization of $\boldsymbol{\gamma}_i^{agg}$ and $\Gamma_i^{agg}$ in DER aggregation. However, the critical, active aggregate flexibility of VPPs depends on specific grid dispatch conditions that are hard to acquire in advance and may vary across reserve scenarios, making prediction inherently challenging. Moreover, the centralized coordination of aggregation and dispatch is impractical due to high computational complexity and privacy concerns related to individual VPPs.

Fortunately, inspired by Lagrangian decomposition [37], we note that the dispatch preferences can be expressed via Lagrange multipliers $\boldsymbol{\lambda}_i$ associated with the equality constraint (15). Here, $\boldsymbol{x}_i = [\boldsymbol{\gamma}_i^{agg}; \Gamma_i^{agg}]$ denotes the optimal shape parameters of the approximate aggregate sets in VPPs, while its duplicate $\boldsymbol{x'}_i$ characterizes the aggregate flexibility set used in the power system dispatch subproblem. In the optimal grid reserve dispatch problem—which minimizes operating costs by leveraging VPP flexibility—this constraint (15) must hold [38].

$$\boldsymbol{x}_i = \boldsymbol{x}'_i : \boldsymbol{\lambda}_i \tag{15}$$

By treating these special Lagrange multipliers $\boldsymbol{\lambda}_i$ as *preference signals*, we propose the distributed aggregation-dispatch coordination framework where the power system shares its Lagrange multipliers, reflecting specific grid dispatch preferences, with the VPPs. These preference signals enable each VPP to incorporate system dispatch preferences into its local aggregation optimization, thereby computing aggregate sets that align with overall system requirements in a distributed, privacy-preserving manner. Consequently, issues related to unpredictability, computational burden, and privacy are effectively mitigated.

Fig. 2 illustrates the proposed aggregation-dispatch coordination framework. During coordination, each VPP locally optimizes its aggregate flexibility set using the received preference signal $\boldsymbol{\lambda}_i$ and satisfying its internal constraints (14.a) -(14.b), then submits its aggregate flexibility set $\mathbb{P}_i^{agg}(\boldsymbol{x}_i)$, parameterized by $\boldsymbol{x}_i$, to the power system. Upon receiving the VPPs' offers, the power system optimizes its reserve dispatch and updates the preference signals $\boldsymbol{\lambda}_i$ to further guide the VPP aggregation. This iterative process continues until convergence, at which point the coordination parameters $\boldsymbol{x}_i$ remain fixed. Detailed formulations are provided below.

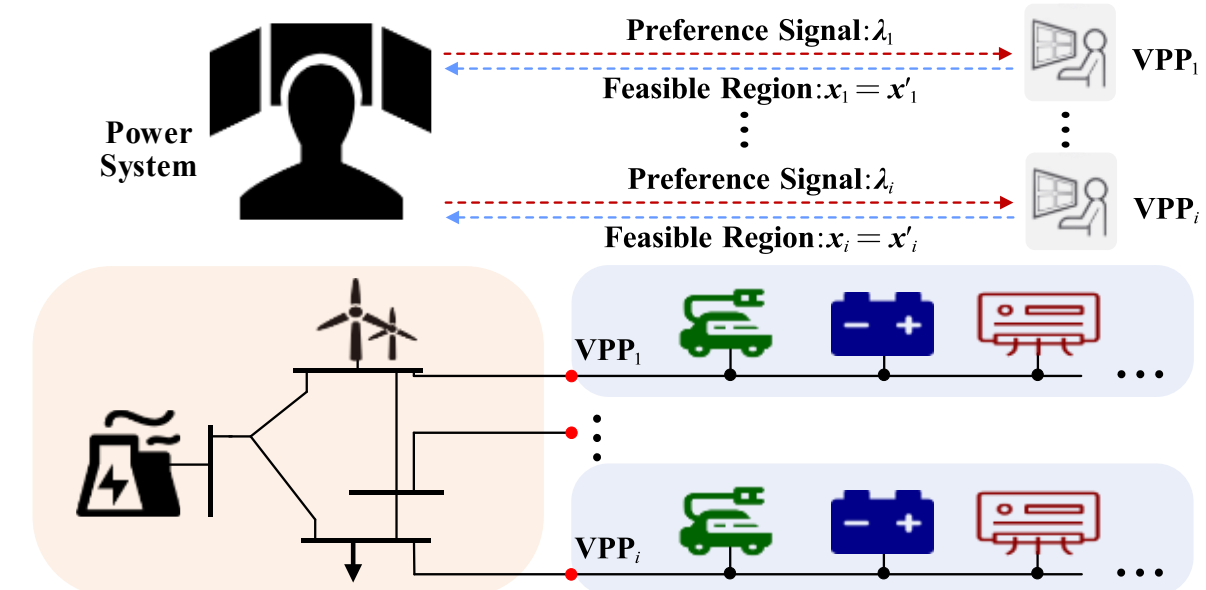

Fig. 2. Aggregation-dispatch coordination framework.

### A. Aggregation Problem of VPPs

In VPP aggregation problem defined in (16), the goal is to optimize the shape parameters, denoted by $\boldsymbol{x}_i$, for the approximate aggregate set based on the preference signals $\boldsymbol{\lambda}_i$ received from the power system, while satisfying the inner approximation constraints (14.a)-(14.b). Here, $\mathcal{L}(\boldsymbol{x}_i, \boldsymbol{x'}_i, \boldsymbol{\lambda}_i)$ is the Lagrangian penalty term.

$$\begin{aligned}&\min_{\boldsymbol{x}_i}\ \mathcal{L}\left(\boldsymbol{x}_i,\boldsymbol{x}_i',\boldsymbol{\lambda}_i\right)\\&\text{s.t. (14.a)-(14.b)}\end{aligned}\quad(16)$$

By applying Lagrangian relaxation to the boundary constraint (15) between the VPPs and the power system, the penalty term $\mathcal{L}(\boldsymbol{x}_i,\boldsymbol{x}'_i,\boldsymbol{\lambda}_i)$ can be defined as follows.

$$\mathcal{L}\left(\boldsymbol{x}_i,\boldsymbol{x}_i',\boldsymbol{\lambda}_i\right)=\boldsymbol{\lambda}^T\left(\boldsymbol{x}-\boldsymbol{x}'\right)+\frac{\rho}{2}\left\|\boldsymbol{x}-\boldsymbol{x}'\right\|_2^2\quad(17)$$

where $\boldsymbol{x}$ represents the collection of all $\boldsymbol{x}_i$ (similarly for $\boldsymbol{x}'$ and $\boldsymbol{\lambda}$); $\rho$ is a predefined positive parameter.

Building on this, we observe that both the objective function and constraints of VPPs are separable from one another, indicating that the vectors $\{\boldsymbol{x}_i\}$ are orthogonal. This yields:

$$\left\|\boldsymbol{x}-\boldsymbol{x}'\right\|_2^2=\left\|\sum_i\left(\boldsymbol{x}_i-\boldsymbol{x}_i'\right)\right\|_2^2=\sum_i\left\|\boldsymbol{x}_i-\boldsymbol{x}_i'\right\|_2^2\quad(18)$$

We then reformulate the Lagrangian term (17) as below:

$$\mathcal{L}\left(\boldsymbol{x}_i,\boldsymbol{x}_i',\boldsymbol{\lambda}_i\right)=\sum_i\boldsymbol{\lambda}_i^T\left(\boldsymbol{x}_i-\boldsymbol{x}'\right)+\frac{\rho}{2}\sum_i\left\|\boldsymbol{x}_i-\boldsymbol{x}_i'\right\|_2^2\quad(19)$$

### *B. Reserve Dispatch Problem of Power System*

The operational model for grid reserve dispatch is formulated as a two-stage stochastic optimization problem, presented in (20.a)-(20.j). The first stage involves the day-ahead reserve capacity dispatch of VPPs and thermal units, while the second stage focuses on scenario-based intra-day regulation, where reserves from both VPPs and thermal units are deployed.

$$\min\ \sum_i\sum_t\left(C_{i,t}^{PS}\right)+\mathcal{L}\left(\boldsymbol{x}_i,\boldsymbol{x}_i',\boldsymbol{\lambda}_i\right)\quad(20.a)$$

$$\begin{aligned}\text{s.t. }C_{i,t}^{PS}&=C_{i,t}^{f}+C_i^{gu}P_{i,t}^{gu}+C_i^{gd}P_{i,t}^{gd}+C_i^{vu}P_{i,t}^{vu}+C_i^{vd}P_{i,t}^{vd}\\&+\pi_s\sum_s\begin{pmatrix}C_i^{gud}P_{i,t,s}^{gud}+C_i^{gdd}P_{i,t,s}^{gdd}+C_i^{vud}P_{i,t,s}^{vud}\\+C_i^{vdd}P_{i,t,s}^{vdd}+C_i^{wc}P_{i,t,s}^{wc}+C_i^{ls}P_{i,t,s}^{ls}\end{pmatrix}\end{aligned}\quad(20.b)$$

$$\sum_i\left(P_{i,t}^{g,fc}+P_{i,t}^{wd,fc}\right)=\sum_i\left(P_{i,t}^{ld,fc}+P_{i,t}^{ref}\right)\quad(20.c)$$

$$-\overline{P}_j^{line}\le\sum_i S_{ji}\left(P_{i,t}^{g,fc}+P_{i,t}^{wd,fc}-P_{i,t}^{ld,fc}-P_{i,t}^{ref}\right)\le\overline{P}_j^{line}\quad(20.d)$$

$$\underline{P}_i^g\le P_{i,t}^{g,fc}\le\overline{P}_i^g,-\overline{\delta}_i^g\le P_{i,t}^{g,fc}-P_{i,t-1}^{g,fc}\le\overline{\delta}_i^g\quad(20.e)$$

$$\sum_i\left(P_{i,t,s}^{g,st}+P_{i,t,s}^{wd,st}-P_{i,t,s}^{wc}\right)=\sum_i\left(P_{i,t,s}^{ld,st}-P_{i,t,s}^{ls}+P_{i,t,s}^{agg}\right)\quad(20.f)$$

$$-\overline{P}_j^{line}\le\sum_i S_{ji}\left(P_{i,t,s}^{g,st}+P_{i,t,s}^{wd,st}-P_{i,t,s}^{wc}-P_{i,t,s}^{ld,st}+P_{i,t,s}^{ls}-P_{i,t,s}^{agg}\right)\le\overline{P}_j^{line}\quad(20.g)$$

$$0\le P_{i,t,s}^{wc}\le P_{i,t,s}^{wd,st},0\le P_{i,t,s}^{ls}\le P_{i,t,s}^{ld,st}\quad(20.h)$$

$$\begin{cases}P_{i,t,s}^{g,st}=P_{i,t}^{g,fc}+P_{i,t,s}^{gud}-P_{i,t,s}^{gdd}\\0\le P_{i,t,s}^{gud}\le P_{i,t}^{gu},0\le P_{i,t,s}^{gud}\le P_{i,t}^{gd}\\\underline{P}_i^g\le P_{i,t,s}^{g,st}\le\overline{P}_i^g,-\overline{\delta}_i^g\le P_{i,t,s}^{g,st}-P_{i,t-1,s}^{g,st}\le\overline{\delta}_i^g\end{cases}\quad(20.i)$$

$$\begin{cases}P_{i,t,s}^{agg}=P_{i,t}^{ref}+P_{i,t,s}^{vud}-P_{i,t,s}^{vdd},\boldsymbol{P}_{i,s}^{agg}\in\mathbb{P}_i^{agg}\left(\boldsymbol{x}_i'\right)\\0\le P_{i,t,s}^{vud}\le P_{i,t}^{vu},0\le P_{i,t,s}^{vud}\le P_{i,t}^{vd}\end{cases}\quad(20.j)$$

(20.a) depicts the objective function, minimizing the total operational cost and the Lagrangian penalty. The costs for the first stage are outlined in the first line of (20.b), which includes the fuel costs of thermal units, the costs of VPP reserve capacity, and the costs of thermal unit reserve capacity. The expected costs for the second stage are detailed in the second line of (20.b) and encompass the upward and downward regulation costs for thermal units and VPPs, as well as penalties for wind curtailment and load shedding. The piecewise linear function for fuel cost $C_{i,t}^f$ is referenced in [39].

In the first stage constraints, (20.c) enforces power balance; (20.d) sets transmission line power flow limits using the DC power flow model, which is widely adopted in day-ahead scheduling [38]; (20.e) establishes power output limits for thermal units. In the second stage constraints, (20.f) encodes power balance; (20.g) sets transmission line power flow limits under the DC power flow model; (20.h) outlines wind curtailment and load shedding limits; (20.i) enforces constraints for thermal units, including up/down reserve deployment, ramping, and generation constraints; (20.j) specifies constraints for VPPs, including up/down reserve deployment and aggregate flexibility set constraints.

### *C. ADMM Solution Algorithm*

*1) Algorithm Description:* The proposed aggregation–dispatch coordination problem is solved using the Alternating Direction Method of Multipliers (ADMM) algorithm [40]. In each iteration $o$, the following three steps are executed sequentially. In the first step (21.a), each VPP optimizes its aggregate flexibility set and determines $\boldsymbol{x}_{i,o}$ given the preference signals $\boldsymbol{\lambda}_{i,o-1}$ and the power system's preferred solution $\boldsymbol{x}'_{i,o-1}$ from the previous iteration. In the second step (21.b), the reserve-dispatch problem is solved, updating the preferred aggregate flexibility sets by $\boldsymbol{x}'_{i,o}$. In the third step (21.c), the preference signals $\boldsymbol{\lambda}_{i,o}$ are updated based on the outcomes of the first two steps. The algorithm proceeds until the stopping criterions (21.d)-(21.f) are satisfied [4], [40], [41]. Here, $R^p/R^d$ are the primal/dual residual during iteration and $\varepsilon$ is the convergence threshold.

$$\begin{aligned}\boldsymbol{x}_{i,o}&=\arg\min_{\boldsymbol{x}_i}\ \mathcal{L}\left(\boldsymbol{x}_i,\boldsymbol{x}_{i,o-1}',\boldsymbol{\lambda}_{i,o-1}\right)\\&\text{s.t. (14.a)-(14.b)}\end{aligned}\quad(21.a)$$

$$\begin{aligned}\boldsymbol{x}_{i,o}'&=\arg\min_{\boldsymbol{x}_i'}\sum_i\sum_t\left(C_{i,t}^{PS}\right)+\mathcal{L}\left(\boldsymbol{x}_{i,o},\boldsymbol{x}_i',\boldsymbol{\lambda}_{i,o-1}\right)\\&\text{s.t. (20.a)-(20.j)}\end{aligned}\quad(21.b)$$

$$\boldsymbol{\lambda}_{i,o}=\boldsymbol{\lambda}_{i,o-1}+\rho\left(\boldsymbol{x}_{i,o}-\boldsymbol{x}_{i,o}'\right)\quad(21.c)$$

$$R_o^p=\left\|\boldsymbol{x}_{i,o}-\boldsymbol{x}_{i,o}'\right\|_2\quad(21.d)$$

$$R_o^d=\rho\left\|\boldsymbol{x}_{i,o}'-\boldsymbol{x}_{i,o-1}'\right\|_2\quad(21.e)$$

$$\max\left\{R_o^p,R_o^d\right\}\le\varepsilon\quad(21.f)$$

**Remark 4** (Parallel-enabled decomposition). Due to the orthogonality of the vectors $\{\boldsymbol{x}_i\}$, (21.a) exhibits a *block-separable* structure concerning each VPP's constraints and objectives. This structure enables the decomposition of problem (21.a) into independent subproblems for each VPP as below:

$$\min_{x_i} \ \boldsymbol{\lambda}_i^T\left(\boldsymbol{x}_i-\boldsymbol{x}'\right)+\frac{\rho}{2}\left\|\boldsymbol{x}_i-\boldsymbol{x}_i'\right\|_2^2 \quad \text{s.t. (14.a)-(14.b)} \tag{22}$$

Thus, the aggregation problems for each VPP can be solved independently and concurrently, enhancing the computational efficiency of the proposed preference-oriented framework. Additionally, the framework safeguards the privacy of both the power system and the VPPs by sharing only the boundary aggregation parameters $\boldsymbol{x}_i$ and the preference signals $\boldsymbol{\lambda}_i$.

**Remark 5** (Convergence behavior) The ADMM algorithm converges for convex problems [40], [42]. In our aggregation–dispatch coordination framework, both the aggregation subproblem for VPPs (22) and the reserve-dispatch subproblem for the power system (21.b) are convex, which guarantees ADMM convergence. However, if our preference-oriented method is applied to more complex, nonconvex power-system dispatch problems, additional modifications would be required to ensure convergence [43]. To further improve convergence performance, we note that existing preference-agnostic aggregation methods, despite not capturing the critical active aggregate flexibility required for optimal reserve dispatch, can provide effective initial guesses to warm-start our preference-oriented framework. This warm-start strategy effectively bridges the gap between conventional preference-agnostic aggregation approaches and our preference-oriented method, which accelerates convergence, reduces the number of iterations, and mitigates potential divergence risks during the aggregation-dispatch coordination process [44].

*2) Algorithm Procedure*: By leveraging the above parallel-enabled property and warm-start acceleration strategy, we are able to efficiently solve the proposed aggregation-dispatch coordination framework. **Algorithm 1** provides a summary of the ADMM algorithm's procedure, as illustrated in Fig. 3.

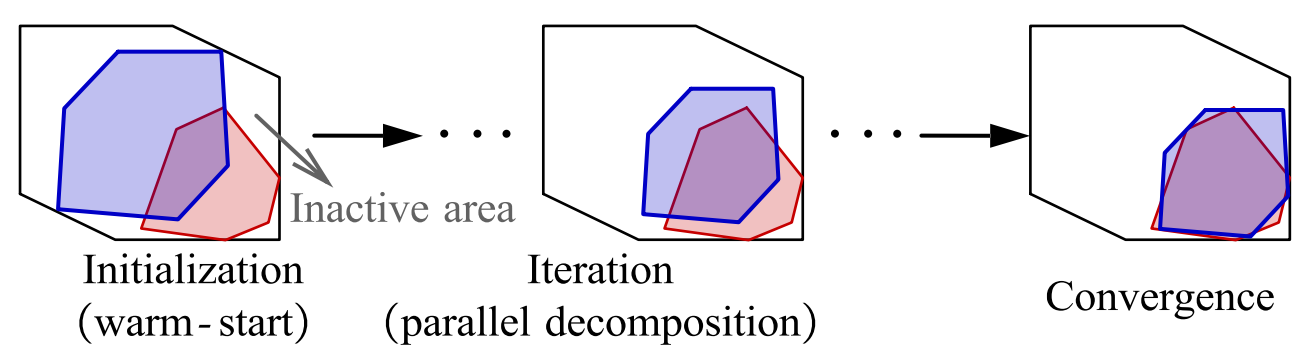

Fig. 3. Illustration of iteration process.

**Algorithm 1.** ADMM algorithm

**Initialize**: Set iterations $o = 0$, initial Lagrangian multiplier $\lambda_{i,0}$, parameter $\boldsymbol{x}'_{i,0}$ , and penalty factor $\rho$.

**Procedure:**

**1:** Solve the aggregation problem (22) in parallel, then solve the grid reserve dispatch problem (21.b) to get the $o-$th iteration solutions for the variables in in (21.c).

**2:** Update Lagrangian multiplier $\boldsymbol{\lambda}_{i,o}$ according to (21.c).

**3:** If (21.d)-(21.f) hold, report finish; otherwise, $o = o + 1$, go to Step 1.

Through iterative optimization, the optimal value of $\boldsymbol{x}_i$ for DER aggregation is determined, ensuring that the derived approximate aggregate set aligns as closely as possible with the critical, active regions favored by optimal reserve dispatch.

### D. Schematic Overview of Proposed Method Implementation

The proposed preference-oriented aggregation method is implemented in the day-ahead grid reserve dispatch to determine the optimal reserve provision of VPPs, consisting of the following two sequential processes:

***Process 1*** *(Preference-Oriented Aggregation)*: VPPs collect the parameters of DERs that they manage and negotiate the aggregate flexibility sets with the power system using the proposed **Algorithm 1**, aligning with grid dispatch preferences.

***Process 2*** *(Grid Reserve Dispatch)*: The power system operator receives the determined aggregate flexibility sets from the VPPs and calculates the optimal reserve dispatch results by solving (20.a)-(20.j) . At this stage, the Lagrangian penalty term $\mathcal{L}(\boldsymbol{x}_i, \boldsymbol{x}'_i, \boldsymbol{\lambda}_i)$ in the objective is no longer considered, and the variables $\boldsymbol{x}'_i$ in the constraints are treated as constants, having been determined during ***Process 1***.

## V. Case Studies

Case studies are conducted on a toy example, the modified 6-bus system, and a real-world regional transmission system to validate the effectiveness of the proposed aggregation method.

The numerical simulations are based on a computer with an Intel i9-14900 CPU 2.20 GHz and 64 GB RAM, running on MATLAB software. The YALMIP optimization toolbox [45] and GUROBI 12.0.1 [36] are used to solve the optimization problems. Due to the challenges that existing aggregation methods face in handling dimensional heterogeneity, we compare the following three cases herein:

**Case 1**: Centralized control without aggregation, where all DER information is comprehensively considered in grid reserve dispatch. This serves as the benchmark for evaluation.

**Case 2 (Proposed matrix transformation technique)**: Preference-agnostic aggregation method based on the proposed matrix transformation, designed to maximize the aggregate set volume as the traditional methods do.

**Case 3 (Proposed matrix transformation technique and preference-oriented insight)**: The proposed preference-oriented aggregation method based on the proposed matrix transformation technique.

### A. An Illustrative Toy Example

We begin by considering a three-period toy example involving two ESSs participating in reserve dispatch. The reserve requirement parameters for three typical scenarios are presented in TABLE I. In these scenarios, positive values indicate that the ESS is required to provide upward reserve (i.e., deliver power externally) at that moment, while negative values indicate power absorption, with each scenario occurring with equal probability. To simplify expression and visualization, charging and discharging power losses are excluded from the case studies [5], [21]. Additionally, ramping limits in (1.d) are disregarded, since DERs are primarily managed by power electronic devices with response times significantly shorter than the grid dispatch intervals [13], [46]. The parameters of ESS#0-2 are given in TABLE II, where ESS#0 is chosen as the base set; ESS#1 provides reserve only in periods 1 and 2, while ESS#2 can provide reserve in all three periods. The ADMM convergence threshold is set to 0.01 [4]. The aggregation and reserve dispatch results are given in Fig. 4 and TABLE III.

TABLE I
RESERVE REQUIREMENT IN TOY EXAMPLE

| Reserve requirement in different scenarios | Period 1 | Period 2 | Period 3 |
|---|---|---|---|
| Scenario 1 (kW) | 10 | 25 | 5 |
| Scenario 2 (kW) | -10 | -15 | -2.5 |
| Scenario 3 (kW) | 30 | -15 | 20 |

TABLE II
DER PARAMETERS IN TOY EXAMPLE

| Parameters | ESS#0 | ESS#1 | ESS#2 |
|---|---|---|---|
| Energy capacity (kWh) | 60 | 40 | 80 |
| Max. charging/discharging power (kW) | 17.5 | 10 | 25 |
| Initial energy (kWh) | 30 | 20 | 40 |
| Energy dissipation rate (p.u.) | 1 | 1 | 1 |
| Available dispatch periods | 1,2,3 | 1,2 | 1,2,3 |

Fig. 4 visualizes the aggregation results for two ESSs with heterogeneous dimensions. The exact aggregate set is constructed by enumerating all extreme points in the aggregate power space [26]. While this is feasible in the toy example, the number of vertices grows exponentially with the period dimension, making such exhaustive computation intractable for practical high-dimensional DER aggregation. In contrast, the proposed inner approximation provides a scalable alternative by adaptively capturing the aggregate flexibility with significantly fewer parameters.

As shown in Fig. 4, although ESS#1 and ESS#2 are represented as polytopes of different dimensions, both **Case 2** and **Case 3**—based on the proposed matrix transformation technique—successfully approximate the overall aggregate flexibility set. This demonstrates our method's effectiveness in overcoming the challenges posed by dimensional heterogeneity in existing aggregation approaches.

Moreover, Fig. 4 demonstrates that although the aggregate set in **Case 3** (preference-oriented) has a smaller volume than that in **Case 2** (preference-agnostic), it still captures the optimal operating points identified in **Case 1**. Specifically, even though the feasible region in **Case 3** occupies only 65.78% of the volume of that in **Case 2**, it yields the same optimal reserve dispatch results as **Case 1** (see TABLE III). This outcome stems from the preference-oriented approach: by integrating reserve dispatch requirements into the aggregation process, **Case 3** targets the critical, active flexibility prioritized in optimal reserve dispatch, thereby yielding high-quality aggregate results. In contrast, **Case 2**'s larger feasible region includes many inactive and ineffective areas, resulting in an expected reserve shortage of 8.44 kWh and underutilizing the aggregate flexibility of the DERs (see TABLE III).

In summary, this toy example demonstrates the effectiveness of our aggregation technique for dimensionally heterogeneous DERs and underscores the value of the preference-oriented insight. A larger approximate aggregate set is not necessarily better; by focusing on capturing only the critical regions, we can yield high-quality aggregate results and achieve reserve dispatch performance comparable to that of the exact set without the inclusion of inactive areas.

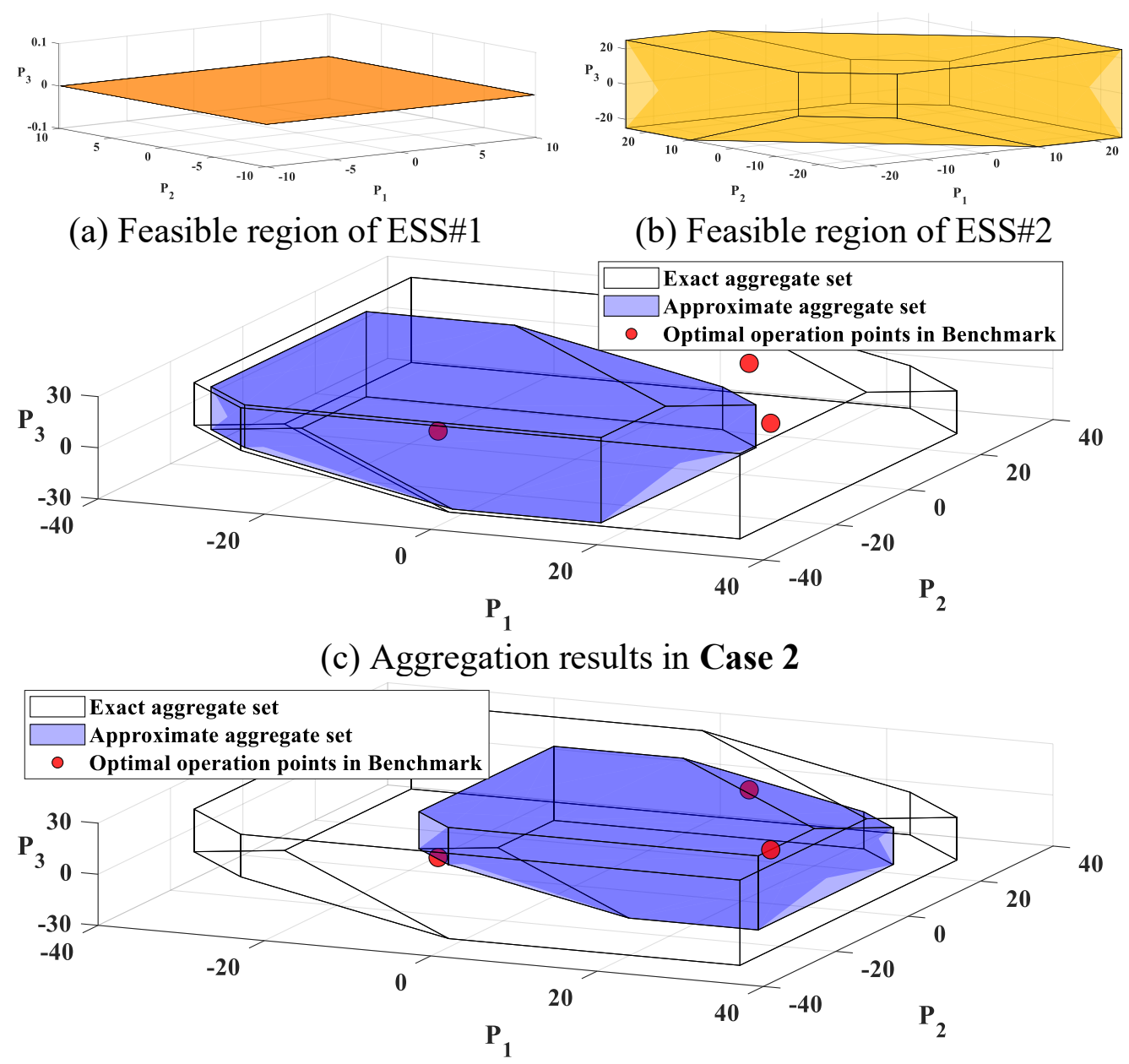

(a) Feasible region of ESS#1 (b) Feasible region of ESS#2

(c) Aggregation results in **Case 2**

(d) Aggregation results in **Case 3**

Fig. 4. Visualization of aggregation results in toy example (*Optimal operation points in Benchmark* refer to the optimal reserve dispatch results of VPP derived from **Case 1** under different reserve requirement scenarios).

TABLE III
AGGREGATION AND DISPATCH RESULTS IN TOY EXAMPLE

| Comparison terms | **Case 1** (exact aggregate set) | **Case 2** | **Case 3** |
|---|---|---|---|
| Aggregate set's volume (p.u.) | 228,541 | 112,776 | 74,186 |
| Expected reserve deployment (kWh) | 44.17 | 35.73 | 44.17 |
| Expected reserve scarcity (kWh) | 0 | 8.44 | 0 |

### *B. Method Validation in Small-Scale System*

We further analyze a modified 6-bus system including three thermal units and an 80 MW wind farm, using data from [47]. The total load is 210 MW, and the dispatch horizon spans 24 hours with a 1-hour granularity per period. Load demand and wind generation power curves are sourced from CAISO, as shown in Appendix D. Ten typical stochastic scenarios of net load are generated and reduced following the methodology in [48], as illustrated in Fig. 5. Penalty costs for load shedding and wind curtailment are set at 1000 $/MWh and 200 $/MWh, respectively [8]. The prices for reserve deployment and reserve capacity for thermal units are 1.3 and 0.4 times their highest incremental price, respectively [8]. A VPP at bus 3 consists of 15 BS clusters, 15 EV clusters, and 15 TCR clusters. For each cluster, the parameters are randomly generated via Monte Carlo sampling (see Appendix D) [13]. The reserve deployment and capacity prices for DERs are set at 13 $/MWh and 4 $/MWh, respectively [49]. The base set is defined as a lossless battery storage model as detailed in Appendix A, with the ratio of its energy capacity to maximum power set at 8.

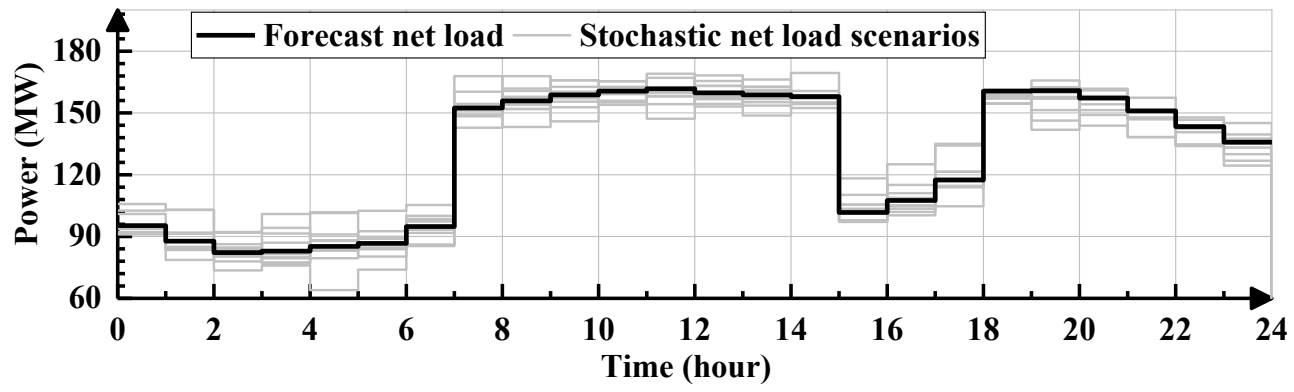

Fig. 5. Stochastic scenarios of net load (*Net load* is defined as the load demand minus the wind power generation).

*1) Convergence Behavior:* We evaluate the convergence performance of ADMM algorithms under two cases, with results shown in Fig. 6:

**Case 3a:** Standard ADMM algorithm without acceleration.

**Case 3b:** ADMM algorithm warm-started from the preference-agnostic solution.

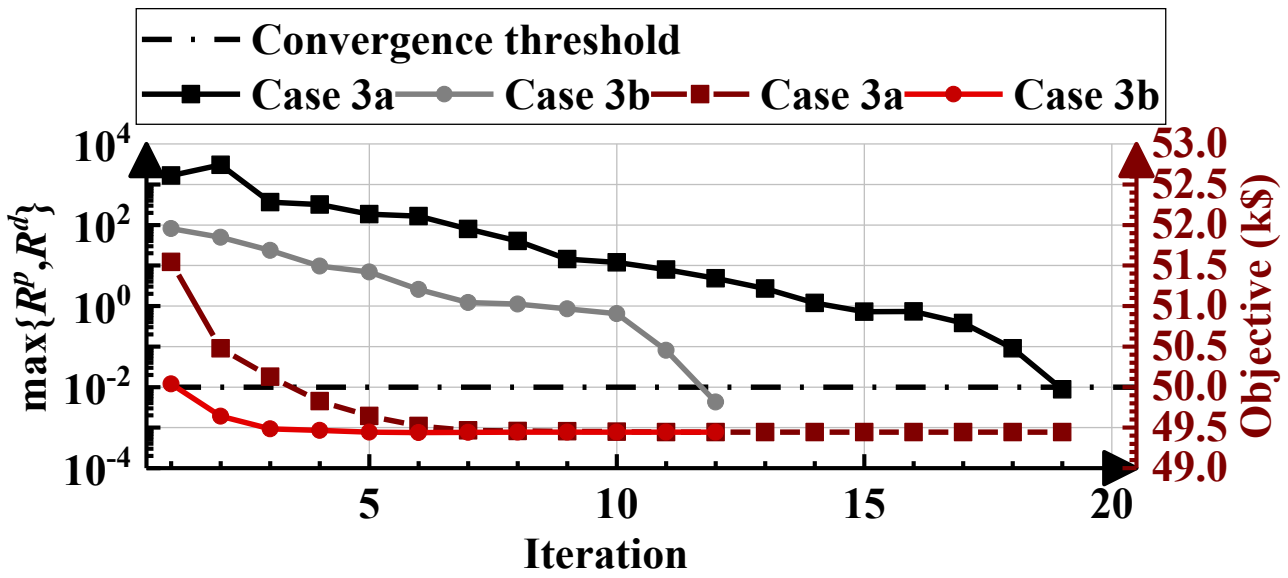

Fig. 6. Convergence behavior in 6-bus system.

Fig. 6 demonstrates that both cases converge gradually. **Case 3a** requires 19 iterations (105 seconds) to reach convergence. In contrast, **Case 3b** uses the preference-agnostic solution as a warm start, achieving an objective function value much closer to the final optimum from the first iteration. This initialization enables **Case 3b** to converge in only 12 iterations (66 seconds). Both cases reach the same final objective value, validating the effectiveness of the proposed warm-start acceleration.

The computational efficiency analysis reveals that each iteration in the preference-oriented aggregation takes approximately 5.5 seconds (about 3.4 seconds for the dispatch subproblem and about 2.1 seconds for the VPP aggregation subproblem). The aggregation of 45 heterogeneous DER clusters requires only 2.1 seconds, demonstrating the efficiency of the proposed linear reformulation. As stated in **Proposition 3**, this reformulation transforms the problem into a linear form that commercial solvers can handle efficiently.

*2) Comparison of Reserve Dispatch Results:* The dispatch results for **Case 1-3** are presented in TABLE IV and Fig. 7.

TABLE IV illustrates that **Case 1**—incorporating all detailed DER information without aggregation—requires the longest dispatch CPU time to reach optimality, making it less feasible for larger-scale systems due to its computational burden. In **Case 2**, where dispatch preferences are disregarded during aggregation, the dispatch results are conservative, with expected DER reserve deployment reaching only 78.67% of that in **Case 1**. This is because maximizing the feasible region without accounting for reserve dispatch needs includes many inactive areas, leading to underutilized DER flexibility and forcing reliance on more expensive thermal units to maintain balance under uncertain wind conditions. As a result, **Case 2** incurs a 21.32% higher total reserve dispatch cost compared to **Case 1**. In contrast, **Case 3** integrates dispatch preferences into the aggregation process, achieving 93.07% of the expected DER reserve deployment relative to **Case 1** with only a 6.30% increase in total reserve dispatch cost. By aligning the aggregated region with dispatch requirements, **Case 3** better meets the power system's needs and delivers improved economic performance over **Case 2**.

TABLE IV
GRID RESERVE DISPATCH RESULTS IN 6-BUS SYSTEM

| Items | Case 1 | Case 2 | Case 3 |
|---|---|---|---|
| First stage energy cost of thermal units (k$) | 45.3 | 45.3 | 45.3 |
| First stage reserve capacity cost of thermal units ($) | 468.4 | 1,621.0 | 895.1 |
| First stage DER reserve capacity cost ($) | 1,770.6 | 1,204.0 | 1,560.5 |
| I: First stage total reserve capacity cost ($) | 2,239.0 | 2,825.0 | 2,455.6 |
| Second stage expected regulation cost of thermal units ($) | 216.0 | 770.08 | 344.9 |
| Second stage expected DER regulation cost ($) | 1,444.3 | 1,135.8 | 1,344.6 |
| Second stage expected wind curtailment cost ($) | 0 | 0 | 0 |
| Second stage expected load shedding cost ($) | 0 | 0 | 0 |
| II: Second stage total expected reserve deployment cost ($) | 1,660.3 | 1,905.8 | 1,689.5 |
| I+II: Total reserve dispatch cost ($) | 3,899.3 | 4,730.8 | 4,145.1 |
| Expected DER reserve deployment (MWh) | 111.1 | 87.4 | 103.4 |
| Dispatch CPU time (s) | 9.78 | 3.45 | 3.38 |

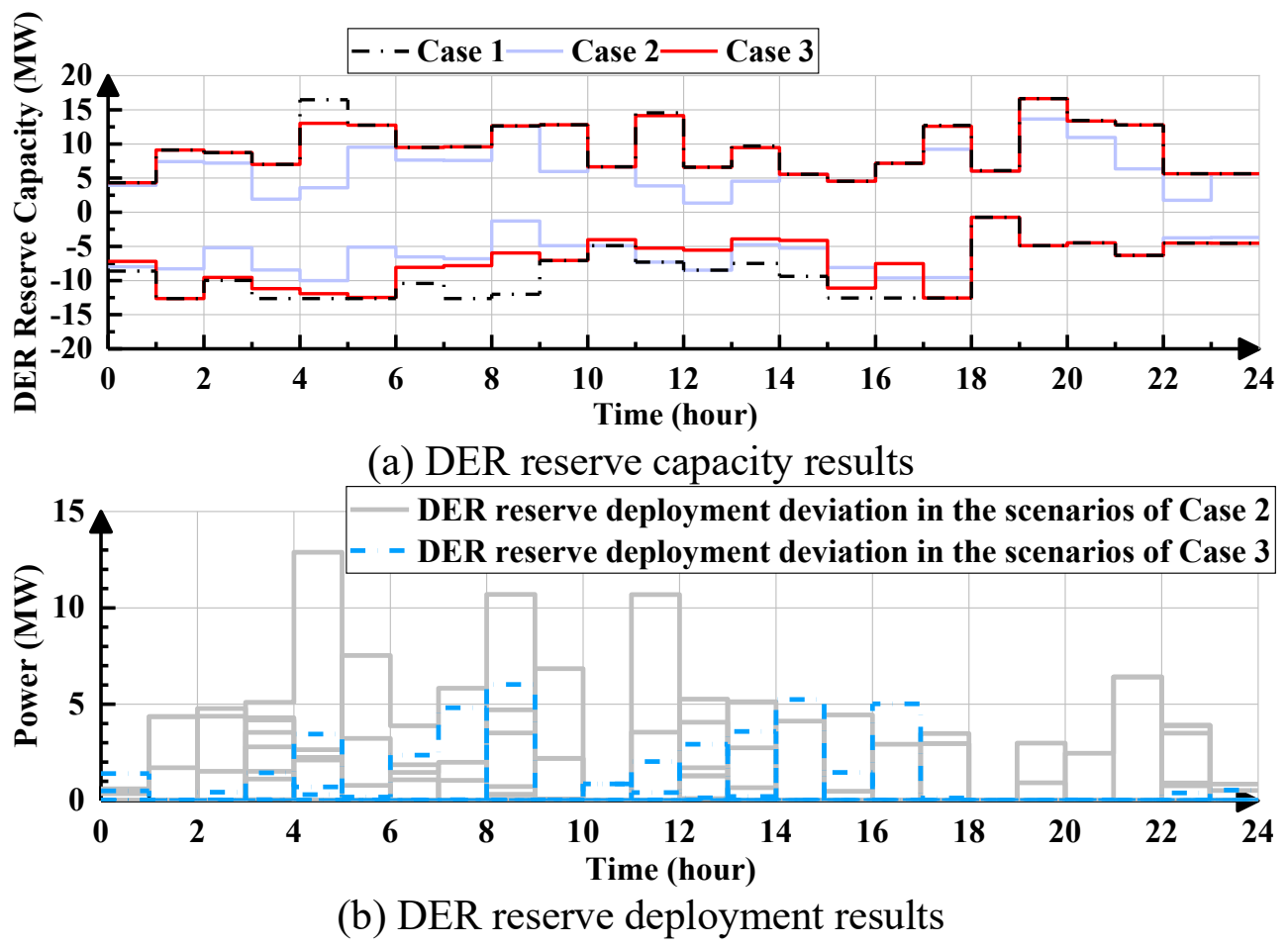

(a) DER reserve capacity results

(b) DER reserve deployment results

Fig. 7. DER reserve dispatch results in 6-bus system (*Reserve deployment deviation* refers to the absolute difference in reserve deployment between the aggregation case (**Case 2** and **Case 3**) and **Case 1** across different scenarios.).

Fig. 7 further compares the dispatch outcomes of **Case 2** and **Case 3**. In Fig. 7(a), **Case 3**'s reserve capacity closely matches that of **Case 1**, whereas **Case 2** fails to fully utilize the DER reserve flexibility. For instance, between 16:00 and 17:00, **Case 2**'s reserve capacity is 16.79 MW—exceeding **Case 3**'s 14.69 MW—but from 17:00 to 18:00, it drops to 18.82 MW, well below **Case 3**'s 25.14 MW. This discrepancy arises from **Case**

**2**'s inability to anticipate the increased reserve demand during 17:00–18:00, resulting in an overall DER reserve capacity that is 29.6% lower than that of **Case 3**. Fig. 7(b) further shows that the reserve deployment deviation of **Case 3** is much smaller compared to **Case 2**, indicating that more potential DER reserve regulation flexibility is exploited in **Case 3** while it is lost in **Case 2**.

Overall, these findings indicate that the proposed preference-oriented aggregation method outperforms traditional preference-agnostic approaches by yielding a higher-quality aggregate result that aligns more closely with dispatch optimality preferences when aggregating heterogeneous DERs. This approach mitigates the conservatism induced by type heterogeneity and leverages more potential DER reserve flexibility to enhance system balance and reduce reserve dispatch costs under stochastic wind conditions.

*3) Sensitivity Analysis:* We first conduct a sensitivity analysis to examine the impact of wind capacity on grid reserve requirements (Fig. 8). As wind capacity increases—and grid reserve needs consequently grow—the advantages of the preference-oriented aggregation (**Case 3**) become more pronounced, as it better leverages DER reserve potential to match system demands. In contrast, the preference-agnostic method (**Case 2**) fails to capture these benefits, underscoring the need to incorporate dispatch preferences—particularly in power systems with high renewable penetration.

Next, Fig. 9 presents a sensitivity analysis on the base set ratio. Across various ratios, **Case 3** consistently yields lower reserve costs than **Case 2**, delivering solutions that are closer to optimal and more economically efficient. However, the choice of base set ratio does affect the reserve cost gap; for example, in **Case 3**, a ratio of 10 results in a gap of only 5.9%, while a ratio of 2 increases the gap to 13.3%. Thus, determining the optimal base set configuration for heterogeneous DERs is an intriguing topic that merits further investigation, although it lies beyond the main scope of this paper.

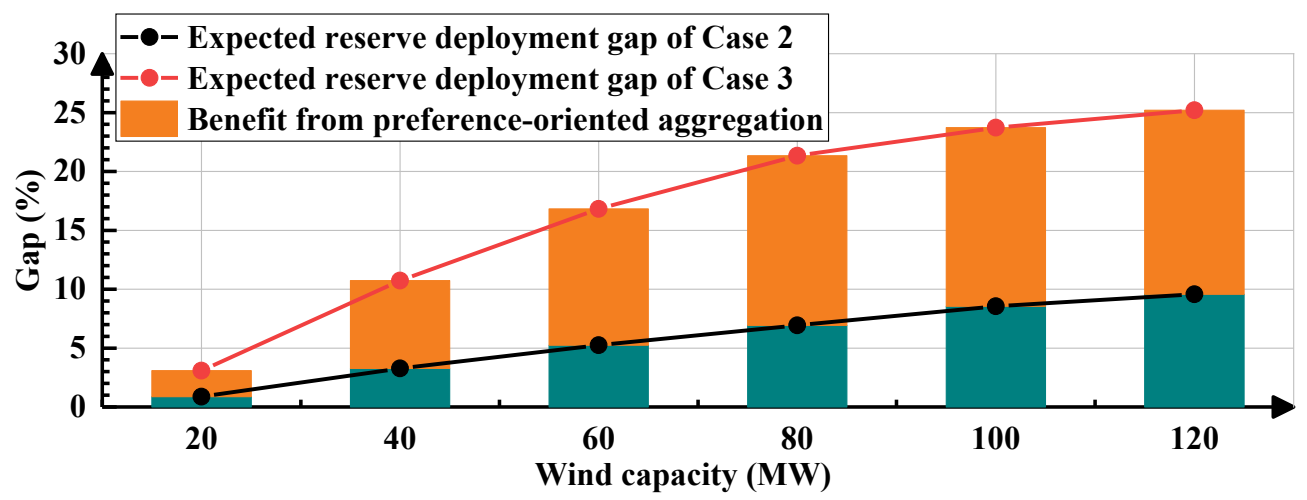

Fig. 8. Sensitivity analysis on wind capacity (*Expected reserve deployment gap* refers to the relative error in expected reserve deployment between the aggregation case and **Case 1**.).

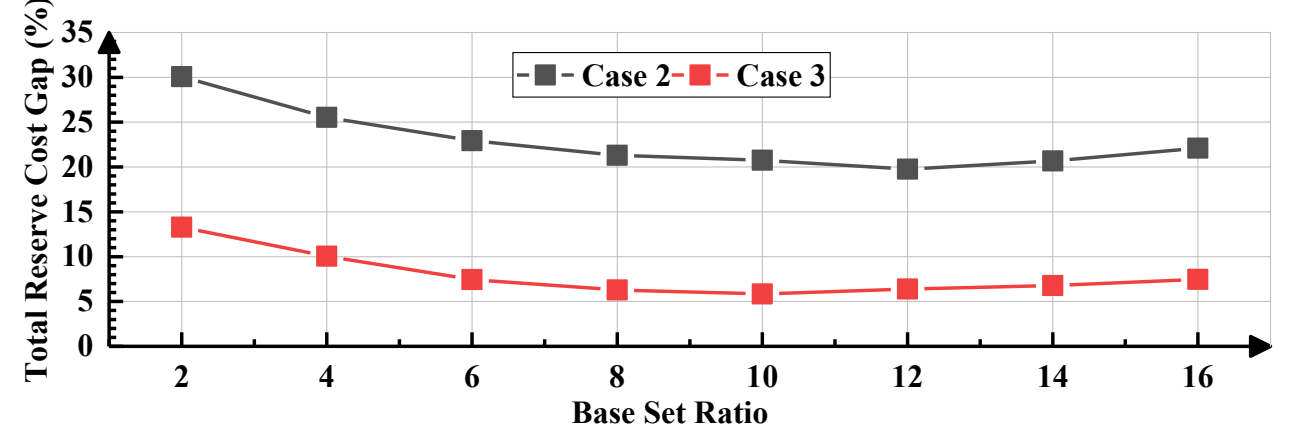

Fig. 9. Sensitivity analysis on base set ratio (*Total reserve cost gap* refers to the relative error in total reserve cost between the aggregation case and **Case 1**; *Base set ratio* refers to the ratio of its energy capacity to maximum power.).

## *C. Scalability Test in Practical-Scale System*

We proceed to validate the scalability of the proposed methods using a practical-scale, real-world regional transmission power system in a province in northern China. Its topology is depicted in Fig. 12 of Appendix D. By 2030, the system is estimated to include 55 500-kV buses and 76 500-kV transmission lines, with a maximum total load of 20,816 MW and line capacities ranging from 400 to 4,000 MW. The network comprises 50 thermal generators with a total capacity of 27,360 MW, complemented by 10,000 MW of wind farms. The net load scenarios, scaled to match the practical-scale system, are analogous to those in Fig. 6. Additionally, 5,400 DER clusters are distributed across 9 VPPs at various nodes, with each VPP comprising 600 DER clusters. To reduce computational complexity during aggregation, each VPP is divided into two levels, forming three groups of 200 clusters. Detailed DER parameters can be found in Appendix D. Here, both the parallel-enabled property among VPPs and the two proposed acceleration strategies are implemented. All other parameters remain consistent with those of the 6-bus system. The results are presented in TABLE V and Fig. 10.

TABLE V
RESERVE DISPATCH RESULTS IN PRACTICAL-SCALE SYSTEM

| Items | **Case 1** | **Case 2** | **Case 3** |
|---|---|---|---|
| First stage energy cost of thermal units (M$) | / | 9.14 | 9.14 |
| First stage reserve capacity cost of thermal units (k$) | / | 291.79 | 167.40 |
| First stage DER reserve capacity cost (k$) | / | 122.55 | 161.23 |
| I: First stage total reserve capacity cost (k$) | / | 414.34 | 328.63 |
| Second stage expected regulation cost of thermal units (k$) | / | 128.85 | 63.80 |
| Second stage expected DER regulation cost (k$) | / | 121.90 | 143.91 |
| Second stage expected wind curtailment cost (k$) | / | 0 | 0 |
| Second stage expected load shedding cost (k$) | / | 0 | 0 |
| II: Second stage total expected reserve deployment cost (k$) | / | 250.75 | 207.71 |
| I+II: Total reserve dispatch cost (k$) | / | 665.09 | 536.34 |
| Expected DER reserve deployment (MWh) | / | $9.4*10^3$ | $11.1*10^3$ |
| Aggregation CPU time | / | 24.66 s | 117.6 min |
| Dispatch CPU time | >1 day | 5.32 min | 5.41 min |

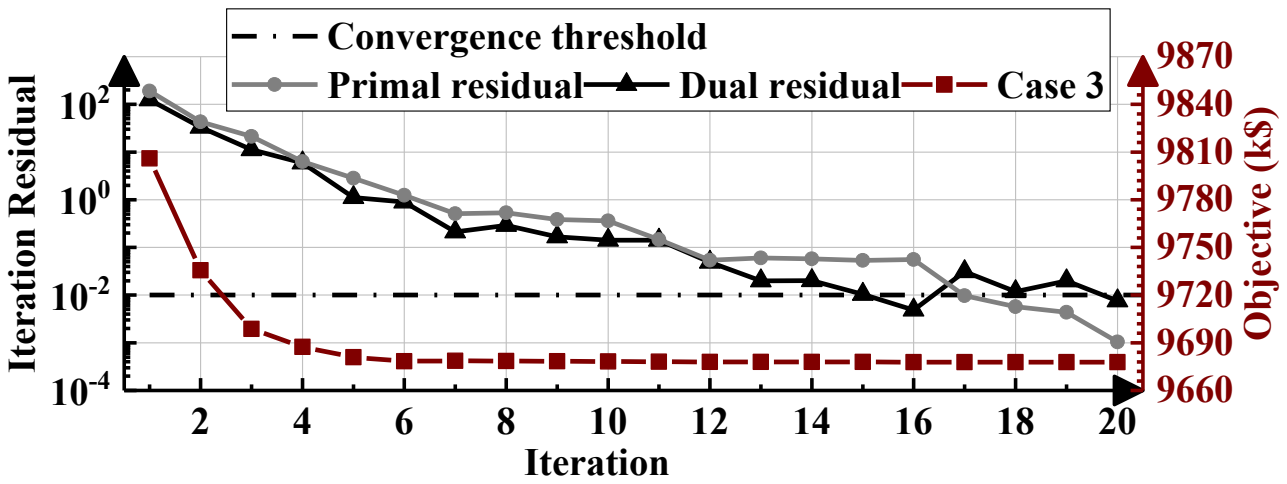

Fig. 10. Convergence behavior in the practical-scale system.

TABLE V shows that **Case 1** fails to achieve optimal reserve dispatch within a day's CPU time due to the heavy computational burden of large-scale DERs. In contrast, **Cases 2** and **Case 3**, which incorporate DER aggregation, solve the reserve dispatch problem within a reasonable time frame. Notably, the preference-oriented aggregation in **Case 3** increases expected DER reserve deployment by 18.1% compared to **Case 2**, fully leveraging DER flexibility and reducing total reserve dispatch costs by 19.4%.

Fig. 10 illustrates the convergence behavior of **Case 3** in the practical-scale system. On average, each iteration takes 5.38 minutes, while the aggregation sub-problems for 9 VPPs are solved in 24.66 seconds via parallel processing (see **Remark 4**). Although each VPP aggregates 600 DER clusters, we employ the grouping strategy from Section III.B to partition them into three parallel groups of 200 clusters. Consequently, each group's aggregation is completed in just 24.66 seconds, underscoring the efficiency of the proposed linear reformulation. Leveraging the warm-start strategy, **Case 3**'s first-iteration total dispatch cost matches that of **Case 2** with comparable CPU time. Subsequent iterations further reduce the cost, reaching the convergence threshold at iteration 20. The total runtime is under two hours, which is acceptable for day-ahead reserve dispatch.

Overall, these results underscore the applicability of the proposed methods for practical-scale power systems.

## VI. Conclusion

This paper presents a preference-oriented aggregation method for heterogeneous DERs in grid reserve dispatch. We develop a matrix transformation-based inner approximation technique to effectively aggregate dimensionally heterogeneous DERs. Furthermore, we propose a preference-oriented aggregation framework that incorporates optimal grid dispatch preferences into the aggregation, enabling the inclusion of the active, critical regions within the aggregate flexibility set.

Validation using a toy example, a modified 6-bus system, and a practical-scale real-world system demonstrates that:

1) Compared to existing polytope-based aggregation techniques, the generalized matrix transformation technique effectively aggregates DERs across different time dimensions, overcoming challenges related to dimensional heterogeneity and facilitating practical implementation.

2) Merely enlarging the volume of the approximate aggregate set does not guarantee enhanced grid reserve dispatch performance. In contrast, by concentrating on the critical, active aggregate regions, our preference-oriented aggregation method produces high-quality aggregate results and significantly reduces the conservatism caused by type heterogeneity. Compared to traditional preference-agnostic methods, the preference-oriented approach identifies more potential DER reserve flexibility, thereby improving system balance and lowering reserve dispatch costs under stochastic wind conditions. These benefits are especially promising for power systems with high renewable penetration, where rising reserve demand calls for unlocking DER flexibility through preference-oriented aggregation.

3) The ADMM algorithm, enhanced by parallel execution and warm-start acceleration strategy, shows computational efficiency in aggregation-dispatch coordination, making it applicable in practical-scale systems.

Future work will concentrate on three main aspects: 1) developing more computationally efficient approaches for preference-oriented aggregation in large-scale power systems; 2) investigating efficient and equitable strategies for power disaggregation; and 3) formulating a method to determine the optimal base set configuration for heterogeneous DERs.

## VII. Appendix

### A. Detailed Models of Distributed Energy Resources

Based on the operational properties of DERs, this paper focuses on three typical types as examples: ESSs, TCRs, and DLs, as introduced below. For detailed derivations and explanations of the DER models, please refer to [13].

*1) Energy Storage Systems:* Considering the charging and discharging schedule along with the state-of-charge limits, the ESS model can be represented as follows:

$$0 \le P_{i,k,t}^{ESS,in} \le \bar{P}_{i,k}^{ESS,in}, \forall t_{i,k}^{ESS,stt} \le t \le t_{i,k}^{ESS,end} \tag{23.a}$$

$$0 \le P_{i,k,t}^{ESS,out} \le \bar{P}_{i,k}^{ESS,out}, \forall t_{i,k}^{ESS,stt} \le t \le t_{i,k}^{ESS,end} \tag{23.b}$$

$$\begin{aligned} -\bar{\delta}_{i,k}^{ESS} \le \left\{ P_{i,k,t-1}^{ESS,in} - P_{i,k,t}^{ESS,in}, P_{i,k,t-1}^{ESS,out} - P_{i,k,t}^{ESS,out} \right\} \le \bar{\delta}_{i,k}^{ESS} \\ , \forall t_{i,k}^{ESS,stt} \le t \le t_{i,k}^{ESS,end} \end{aligned} \tag{23.c}$$

$$\begin{aligned} E_{i,k,t}^{ESS} = \theta_{i,k}^{ESS} E_{i,k,t-1}^{ESS} + \left( \kappa_{i,k}^{ESS,in} P_{i,k,t}^{ESS,in} - P_{i,k,t}^{ESS,out} / \kappa_{i,k}^{ESS,out} \right) \Delta t \\ , \forall t_{i,k}^{ESS,stt} \le t \le t_{i,k}^{ESS,end} \end{aligned} \tag{23.d}$$

$$\underline{E}_{i,k}^{ESS} \le E_{i,k,t}^{ESS} \le \bar{E}_{i,k}^{ESS}, \forall t_{i,k}^{ESS,stt} \le t \le t_{i,k}^{ESS,end} \tag{23.e}$$

where $P_{i,k,t}^{ESS,in}$, $P_{i,k,t}^{ESS,out}$, and $E_{i,k,t}^{ESS}$ are the charging power, discharging power and storage energy, respectively; $\bar{\delta}_{i,k}^{ESS}$ is the ramping limit; $\theta_{i,k}^{ESS}$ is the energy dissipation rate; $\bar{P}_{i,k}^{ESS,in}/\bar{P}_{i,k}^{ESS,out}$ is the maximum charging/ discharging power; $\bar{E}_{i,k}^{ESS}$ / $\underline{E}_{i,k}^{ESS}$ is the maximum/ minimum storage energy constrained by the state-of-charge; $\kappa_{i,k}^{ESS,in}$ / $\kappa_{i,k}^{ESS,out}$ is the efficiency of charging/ discharging power; $t_{i,k}^{ESS,stt}/t_{i,k}^{ESS,end}$ is the available start/end period.

*2) Thermal Controllable Residents:* Accounting for thermal inertia, the dynamic characteristics of TCRs can be represented by a derivative function, which is then converted into the following discrete form:

$$\underline{P}_{i,k}^{TCR,in} \le P_{i,k,t}^{TCR,in} \le \bar{P}_{i,k}^{TCR,in}, \forall t_{i,k}^{TCR,stt} \le t \le t_{i,k}^{TCR,end} \tag{24.a}$$

$$\underline{E}_{i,k}^{TCR} \le E_{i,k,t}^{TCR} \le \bar{E}_{i,k}^{TCR}, \forall t_{i,k}^{TCR,stt} \le t \le t_{i,k}^{TCR,end} \tag{24.b}$$

$$\begin{aligned} -\bar{\delta}_{i,k}^{TCR} \le \left\{ P_{i,k,t-1}^{TCR,in} - P_{i,k,t}^{TCR,in}, P_{i,k,t-1}^{TCR,out} - P_{i,k,t}^{TCR,out} \right\} \le \bar{\delta}_{i,k}^{TCR} \\ , \forall t_{i,k}^{TCR,stt} \le t \le t_{i,k}^{TCR,end} \end{aligned} \tag{24.c}$$

$$\begin{aligned} E_{i,k,t}^{TCR} = \theta_{i,k}^{TCR} E_{i,k,t-1}^{TCR} + \eta_{i,k}^{TCR} P_{i,k,t}^{TCR,in} \Delta t + \omega_{i,k}^{TCR} w_{i,k,t}^{TCR} \\ , \forall t_{i,k}^{TCR,stt} \le t \le t_{i,k}^{TCR,end} \end{aligned} \tag{24.d}$$

where $P_{i,k,t}^{TCR,in}$, and $E_{i,k,t}^{TCR}$ are the power consumption and indoor temperature, respectively; $\bar{P}_{i,k}^{TCR,in}$ / $\underline{P}_{i,k}^{TCR,in}$ is the maximum/ minimum power consumption; $\bar{E}_{i,k}^{TCR}$ / $\underline{E}_{i,k}^{TCR}$ is the maximum/

minimum indoor temperature; $\bar{\delta}_{i,k}^{TCR}$ is the ramping limit; $w_{i,k,t}^{TCR}$ is the ambient temperature; $\theta_{i,k}^{TCR}$ is the dissipation rate; $\eta_{i,k}^{TCR}$ is the conversion coefficient from the active power to temperature; $\omega_{i,k}^{TCR} = 1 - \theta_{i,k}^{TCR}$ is the impact factor of ambient temperature [50]; $t_{i,k}^{DL,stt}/t_{i,k}^{DL,end}$ is the available start/end period.

*3) Deferrable Loads:* The charging load of plug-in EVs is a common example of DLs. This type of DER must meet its energy requirement within a specified available timeframe, which can be expressed as follows:

$$\underline{P}_{i,k}^{DL,in} \le P_{i,k,t}^{DL,in} \le \bar{P}_{i,k}^{DL,in}, \forall t_{i,k}^{DL,stt} \le t \le t_{i,k}^{DL,end} \tag{25.a}$$

$$0 \le E_{i,k,t}^{DL} \le \bar{E}_{i,k}^{DL}, \forall t_{i,k}^{DL,stt} \le t \le t_{i,k}^{DL,end} \tag{25.b}$$

$$-\bar{\delta}_{i,k}^{DL} \le \left\{ P_{i,k,t-1}^{DL,in} - P_{i,k,t}^{DL,in}, P_{i,k,t-1}^{DL,out} - P_{i,k,t}^{DL,out} \right\} \le \bar{\delta}_{i,k}^{DL}, \forall t_{i,k}^{DL,stt} \le t \le t_{i,k}^{DL,end} \tag{25.c}$$

$$E_{i,k,t}^{DL} = \theta_{i,k}^{DL} E_{i,k,t-1}^{DL} + \eta_{i,k}^{DL} P_{i,k,t}^{DL} \Delta t, \forall t_{i,k}^{DL,stt} \le t \le t_{i,k}^{DL,end} \tag{25.d}$$

$$E_{i,k,t}^{DL} \ge E_{i,k}^{DL,rqr}, \text{for } t = t_{i,k}^{DL,end} \tag{25.e}$$

where $P_{i,k,t}^{DL,in}$, and $E_{i,k,t}^{DL}$ are the electric power input and residual energy, respectively; $\bar{P}_{i,k}^{DL,in}/\underline{P}_{i,k}^{DL,in}$ is the maximum/ minimum charging power; $\bar{E}_{i,n}^{DL}$ is the rated energy capacity; $\bar{\delta}_{i,k}^{DL}$ is the ramping limit; $\theta_{i,k}^{DL}$ is the dissipation rate; $\eta_{i,k}^{DL}$ is the conversion coefficient; $E_{i,k}^{DL,rqr}$ is the charging energy requirement; $t_{i,k}^{DL,stt}/t_{i,k}^{DL,end}$ is the available start/end period.

### B. Derivation of DER Coefficient Matrices

The derivation of the coefficient matrix involves two key steps. First, QR decomposition and Gaussian elimination are used to remove equality constraints, eliminating the variables $P_{i,k,t}^{DER,in}$ and $P_{i,k,t}^{DER,out}$. Next, Fourier-Motzkin elimination (FME) and redundancy identification techniques are applied to remove inequality constraints, eliminating the variable $E_{i,k,t}^{DER}$ and leaving only the inequality constraints in terms of $P_{i,k,t}^{DER}$.

*1) Dealing with Equality Constraints*: The equality constraints (1.e)-(1.f) are first recast into the following compact form, where $\boldsymbol{H}_{i,k}^{eq}/\boldsymbol{h}_{i,k}^{eq}$ is the known coefficient matrix/vector derived from (1.e)-(1.f).

$$\boldsymbol{H}_{i,k}^{eq} \left[ \boldsymbol{P}_{i,k}^{DER,in}; \boldsymbol{P}_{i,k}^{DER,out}; \boldsymbol{P}_{i,k}^{DER}; \boldsymbol{E}_{i,k}^{DER} \right] = \boldsymbol{h}_{i,k}^{eq} \tag{26.a}$$

The QR decomposition is applied to factorize $\boldsymbol{H}_{i,k}^{eq}$, yielding (26.b)-(26.c), where $\boldsymbol{Q}_{i,k}^{eq}$ is an orthogonal matrix and $\boldsymbol{R}_{i,k}^{eq,1}$ is an invertible upper triangular matrix:

$$\boldsymbol{H}_{i,k}^{eq} = \boldsymbol{Q}_{i,k}^{eq} \left[ \boldsymbol{R}_{i,k}^{eq,1}, \boldsymbol{R}_{i,k}^{eq,2} \right] \tag{26.b}$$

$$\begin{aligned} &\boldsymbol{R}_{i,k}^{eq,1} \left[ \boldsymbol{P}_{i,k}^{DER,in}; \boldsymbol{P}_{i,k}^{DER,out} \right] + \boldsymbol{R}_{i,k}^{eq,2} \left[ \boldsymbol{P}_{i,k}^{DER}; \boldsymbol{E}_{i,k}^{DER} \right] = \left( \boldsymbol{Q}_{i,k}^{eq} \right)^{\mathrm{T}} \boldsymbol{h}_{i,k}^{eq} \Rightarrow \\ &\left[ \boldsymbol{P}_{i,k}^{DER,in}; \boldsymbol{P}_{i,k}^{DER,out} \right] = \left( \boldsymbol{R}_{i,k}^{eq,1} \right)^{-1} \left( \left( \boldsymbol{Q}_{i,k}^{eq} \right)^{\mathrm{T}} \boldsymbol{h}_{i,k}^{eq} - \boldsymbol{R}_{i,k}^{eq,2} \left[ \boldsymbol{P}_{i,k}^{DER}; \boldsymbol{E}_{i,k}^{DER} \right] \right) \end{aligned} \tag{26.c}$$

Next, we apply Gaussian elimination to substitute equation (26.c) into constraints (1.a)-(1.b), thereby eliminating the variables $\boldsymbol{P}_{i,k}^{DER,in}$ and $\boldsymbol{P}_{i,k}^{DER,out}$. We can then recast the original set (1.a)-(1.f) into the following full-dimensional set $\mathbb{U}_{i,k}^{DER}$, characterized by linear inequalities that involve only the variables $\boldsymbol{P}_{i,k}^{DER}$ and $\boldsymbol{E}_{i,k}^{DER}$ where $\boldsymbol{H}_{i,k}^{ineq}/\boldsymbol{h}_{i,k}^{ineq}$ is the known coefficient matrix/vector derived from the combination of (26.c) and (1.a)-(1.d).

$$\mathbb{U}_{i,k}^{DER} = \left\{ \boldsymbol{H}_{i,k}^{ineq} \left[ \boldsymbol{P}_{i,k}^{DER}; \boldsymbol{E}_{i,k}^{DER} \right] \le \boldsymbol{h}_{i,k}^{ineq} \right\} \tag{26.d}$$

*2) Dealing with Inequality Constraints*: The approach outlined in [17] is employed to combine FME and redundancy constraint identification techniques for eliminating extraneous variables $\boldsymbol{E}_{i,k}^{DER}$ in (26.d). The key strategy involves first using FME to remove a target variable, followed by applying redundancy identification to eliminate any redundant constraints in the resulting set. This process is iteratively continued to remove the next variable until all extraneous variables are eliminated. Finally, the original DER set (1) can be recast as the full-dimensional and non-redundant set (2).

Note that a natural concern about FME is its exponential complexity, primarily because redundant constraints accumulate during the process. However, we emphasize that the combined approach of FME and redundancy constraint identification remains highly applicable to the DER models in this paper for two main reasons: first, as noted in reference [17], embedding redundancy constraint identification at each step of the FME process can effectively reduce its complexity; second, and more importantly, the DER model presented in this paper allows for parallel elimination, as each DER model operates independently. Therefore, the complexity of FME does not significantly increase with the scale of DERs.

### C. Proof of **Proposition 3**

The set inclusion $\boldsymbol{\gamma}_i^{agg} + \Gamma_i^{agg} \mathbb{U}_i^{base} \subseteq \boldsymbol{\Gamma}_i^{aff} \mathbb{U}_i^{aff}$ holds if and only if for any direction $\boldsymbol{c} \in \mathbb{R}^{N^T}$, (27.a) holds:

$$\max_{\boldsymbol{P}_i^{agg} \in \mathbb{U}_i^{base}} \boldsymbol{c}^{\mathrm{T}} (\boldsymbol{\gamma}_i^{agg} + \Gamma_i^{agg} \boldsymbol{I}_i^{agg} \boldsymbol{P}_i^{agg}) \le \max_{\boldsymbol{P}_i^{aff} \in \mathbb{U}_i^{aff}} \boldsymbol{c}^{\mathrm{T}} \boldsymbol{\Gamma}_i^{aff} \boldsymbol{P}_i^{aff} \tag{27.a}$$

From the strong duality of linear programs, we can recast the right-hand side of (27.a) as its dual form as shown in (27.b):

$$\max_{\boldsymbol{P}_i^{agg} \in \mathbb{U}_i^{base}} \boldsymbol{c}^{\mathrm{T}} (\boldsymbol{\gamma}_i^{agg} + \Gamma_i^{agg} \boldsymbol{I}_i^{agg} \boldsymbol{P}_i^{agg}) \le \min_{\boldsymbol{u} \in \mathbb{R}_+^{\mathrm{row}\left(\boldsymbol{H}_i^{aff}\right)}, \boldsymbol{u}^{\mathrm{T}} \boldsymbol{H}_i^{aff} = \boldsymbol{c}^{\mathrm{T}} \boldsymbol{\Gamma}_i^{aff}} \boldsymbol{u}^{\mathrm{T}} \boldsymbol{h}_i^{aff} \tag{27.b}$$

Since the minimum of the right-hand side set is higher than the maximum of the left-hand side set, it indicates that every element in the right-hand side set is greater than those in the left-hand side set. Thus, the following relation holds for $\forall \boldsymbol{c} \in \mathbb{R}^{N^T}, \forall \boldsymbol{u} \in \mathbb{R}_+^{\mathrm{row}(\boldsymbol{H}_i^{aff})}, \boldsymbol{u}^{\mathrm{T}} \mathbf{H}_i^{aff} = \boldsymbol{c}^{\mathrm{T}} \boldsymbol{\Gamma}_i^{aff}, \forall \boldsymbol{P}_i^{agg} \in \mathbb{U}_i^{base}$:

$$\boldsymbol{c}^{\mathrm{T}} (\boldsymbol{\gamma}_i^{agg} + \Gamma_i^{agg} \boldsymbol{I}_i^{agg} \boldsymbol{P}_i^{agg}) \le \boldsymbol{u}^{\mathrm{T}} \boldsymbol{h}_i^{aff} \tag{27.c}$$

Next, we prove the existence of (14.a) and (14.b) individually.

*1) Proof of (14.a)*: Since $\boldsymbol{\gamma}_i^{agg} + \Gamma_i^{agg} \mathbb{U}_i^{base} \subseteq \boldsymbol{\Gamma}_i^{aff} \mathbb{U}_i^{aff}$, there exists $\boldsymbol{P}_i^{agg,0} \in \mathbb{U}_i^{base}$ and $\boldsymbol{P}_i^{aff,0} \in \mathbb{U}_i^{aff}$ such that:

$$\boldsymbol{\gamma}_i^{agg} + \Gamma_i^{agg} \boldsymbol{I}_i^{agg} \boldsymbol{P}_i^{agg,0} = \boldsymbol{\Gamma}_i^{aff} \boldsymbol{P}_i^{aff,0} \tag{27.d}$$

Since $\mathbb{U}_i^{base}$ is full-dimensional, there exists $\epsilon_i \ge 0$, and a $N^{\mathrm{T}}$-dimensional unit box $\mathbb{B}_i$ such that $\boldsymbol{P}_i^{agg,0} + \epsilon_i \mathbb{B}_i \subset \mathbb{U}_i^{base}$ holds [30]. Further, we can derive:

$$\boldsymbol{\gamma}_i^{agg} + \Gamma_i^{agg} \boldsymbol{I}_i^{agg} \left( \boldsymbol{P}_i^{agg,0} + \epsilon_i \mathbb{B}_i \right) \subset \mathbb{U}_i^{aff} \tag{27.e}$$

Thus, there exists $\boldsymbol{\psi}_i \in \mathbb{R}^{N_i^{AP,total}}$ such that:

$$\boldsymbol{\gamma}_i^{agg} + \Gamma_i^{agg} \boldsymbol{I}_i^{agg} \left( \boldsymbol{P}_i^{agg,0} + \epsilon_i \boldsymbol{e}_i \right) = \boldsymbol{\Gamma}_i^{aff} \left( \boldsymbol{P}_i^{aff} + \boldsymbol{\psi}_i \right) \quad (27.f)$$

where $\boldsymbol{e}_i$ is the unit vector in the Cartesian directions.

Thus, we have $\Gamma_i^{agg} \boldsymbol{I}_i^{agg} \boldsymbol{e}_i = \boldsymbol{\Gamma}_i^{aff} \boldsymbol{\psi}_i$, implying that $\Gamma_i^{agg} \boldsymbol{I}_i^{agg} \in \text{range}(\boldsymbol{\Gamma}_i^{aff})$. Here, given matrix $\boldsymbol{A}$, we denote $\text{range}(\boldsymbol{A})$ as its column-space. Then, we denote the space transformation matrix as $\boldsymbol{G}_i^{aux} \in \mathbb{R}^{N_i^{AP,total} \times N^T}$, thereby deriving $\Gamma_i^{agg} \boldsymbol{I}_i^{agg} = \boldsymbol{\Gamma}_i^{aff} \boldsymbol{G}_i^{aux}$.

Also, we have $-\boldsymbol{\gamma}_i^{agg} = \Gamma_i^{agg} \boldsymbol{I}_i^{agg} \boldsymbol{P}_i^{agg,0} - \boldsymbol{\Gamma}_i^{aff} \boldsymbol{P}_i^{aff}$, implying that $-\boldsymbol{\gamma}_i^{agg} \in \text{range}(\boldsymbol{\Gamma}_i^{aff})$. Then, we denote the space transformation matrix as $\boldsymbol{\beta}_i^{aux} \in \mathbb{R}^{N_i^{AP,total}}$, thereby deriving $-\boldsymbol{\gamma}_i^{agg} = \boldsymbol{\Gamma}_i^{aff} \boldsymbol{\beta}_i^{aux}$.

This completes the proof of (14.a).

*2) Proof of (14.b)*: According to (14.a), and $\boldsymbol{u}^{\mathrm{T}} \boldsymbol{H}_i^{aff} = \boldsymbol{c}^{\mathrm{T}} \boldsymbol{\Gamma}_i^{aff}$ in(27.b), we can recast (27.c) as:

$$\boldsymbol{u}^{\mathrm{T}} \left( \boldsymbol{h}_i^{aff} + \boldsymbol{H}_i^{aff} \boldsymbol{\beta}_i^{aux} - \boldsymbol{H}_i^{aff} \boldsymbol{G}_i^{aux} \boldsymbol{P}_i^{agg} \right) \geq 0 \quad (27.g)$$

To efficiently find the minimum in (27.g), we provide a conservative approximation by dropping constraint $\boldsymbol{u}^{\mathrm{T}} \boldsymbol{H}_i^{aff} = \boldsymbol{c}^{\mathrm{T}} \boldsymbol{\Gamma}_i^{aff}$. Then, $\boldsymbol{u}$ is only constrained to be non-negative, implying that $\boldsymbol{h}_i^{aff} + \boldsymbol{H}_i^{aff} \boldsymbol{\beta}_i^{aux} - \boldsymbol{H}_i^{aff} \boldsymbol{G}_i^{aux} \boldsymbol{P}_i^{agg} \geq 0, \forall \boldsymbol{P}_i^{agg} \in \mathbb{U}_i^{base}$. Then, we form a polytope $\mathbb{Q}_i = \left\{ \boldsymbol{P}_i^{agg} \in \mathbb{R}^{N^T} \mid \boldsymbol{H}_i^{aff} \boldsymbol{G}_i^{aux} \boldsymbol{P}_i^{agg} \leq \boldsymbol{h}_i^{aff} + \boldsymbol{H}_i^{aff} \boldsymbol{\beta}_i^{aux} \right\}$, implying that $\mathbb{Q}_i \subseteq \mathbb{P}_i^{agg}$. According to [34], the polytope containment relationship, i.e., $\mathbb{Q}_i \subseteq \mathbb{P}_i^{agg}$, holds if there exists a auxiliary variable matrix $\boldsymbol{\Lambda}_i^{aux} \in \mathbb{R}_+^{\text{row}(\boldsymbol{H}_i^{aff}) \times \text{row}(\boldsymbol{H}_i^{base})}$ such that $\boldsymbol{\Lambda}_i^{aux} \boldsymbol{H}_i^{base} = \boldsymbol{H}_i^{aff} \boldsymbol{G}_i^{aux}, \boldsymbol{\Lambda}_i^{aux} \boldsymbol{h}_i^{base} \leq \boldsymbol{h}_i^{aff} + \boldsymbol{H}_i^{aff} \boldsymbol{\beta}_i^{aux}$.

This completes the proof of (14.b). ■

### *D. Parameters in Case Studies*

TABLE VI
PARAMETER PROBABILITY DISTRIBUTION OF ESS CLUSTERS

| Parameters | Distr. | Mean | Deviation | Min | Max |
|---|---|---|---|---|---|
| Energy capacity (MWh) | TGD | 2.5 | 0.5 | 1 | 4 |
| Max. charging/discharging power (MW) | TGD | 0.64 | 0.32 | 0.16 | 1.25 |
| Initial energy (MWh) | TGD | 0.5 | 0.22 | 0.1 | 0.8 |
| Energy dissipation rate (p.u.) | UD | -- | -- | 0.98 | 1.0 |
| Available start time | UD | -- | -- | 1 | 15 |
| Available end time | UD | -- | -- | 10 | 24 |

*TGD: truncated Gaussian distributions. UD: uniformly distributed.

TABLE VII
PARAMETER PROBABILITY DISTRIBUTION OF PLUG-IN EV CLUSTERS

| Parameters | Distr. | Mean | Deviation | Min | Max |
|---|---|---|---|---|---|
| Energy requirement (MWh) | TGD | 2 | 0.5 | 1 | 2.5 |
| Max. charging power (MW) | TGD | 0.6 | 0.32 | 0.25 | 1 |
| Initial energy (MWh) | TGD | 0.75 | 0.25 | 0.3 | 1.75 |
| Energy dissipation rate (p.u.) | UD | -- | -- | 0.98 | 1.0 |
| Conversion coefficient (MWh/MW) | UD | -- | -- | 0.9 | 1.0 |
| Available start time | UD | -- | -- | 1 | 15 |
| Available end time | UD | -- | -- | 11 | 24 |

TABLE VIII
PARAMETER PROBABILITY DISTRIBUTION OF TCR CLUSTERS

| Parameters | Distr. | Mean | Deviation | Min | Max |
|---|---|---|---|---|---|
| Max. indoor temperature (°C) | TGD | 24 | 0.5 | 22.5 | 25.5 |
| Min. indoor temperature (°C) | TGD | 19 | 0.5 | 20.5 | 17.5 |
| Initial indoor temperature (°C) | TGD | 21 | 0.3 | 20 | 22 |
| Max power consumption (MW) | TGD | 1.5 | 0.25 | 1 | 2 |
| Temperature dissipation rate (p.u.) | UD | -- | -- | 0.94 | 0.98 |
| Conversion coefficient (°C/MW) | UD | -- | -- | 1.6 | 2 |
| Available start time | UD | -- | -- | 1 | 16 |
| Available end time | UD | -- | -- | 12 | 24 |

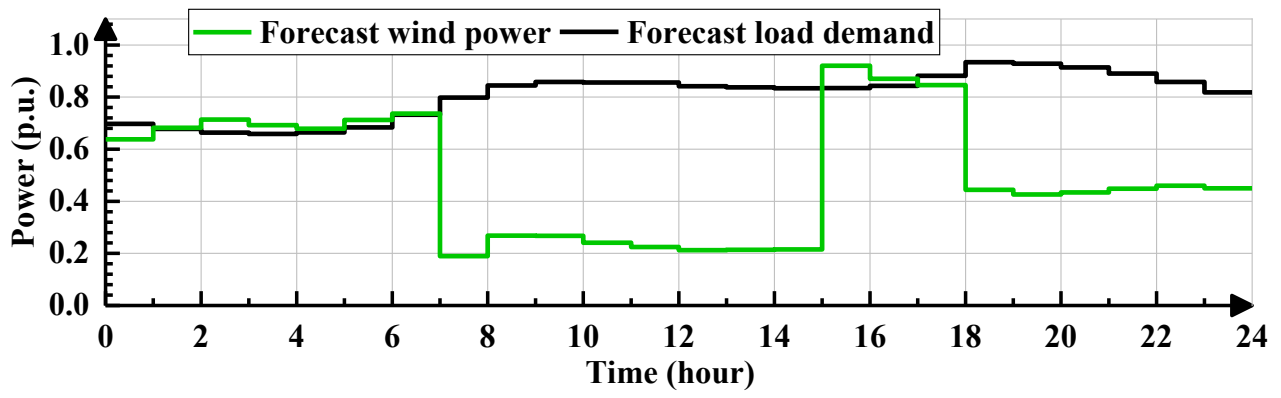

Fig. 11. Load demand and wind power profiles.

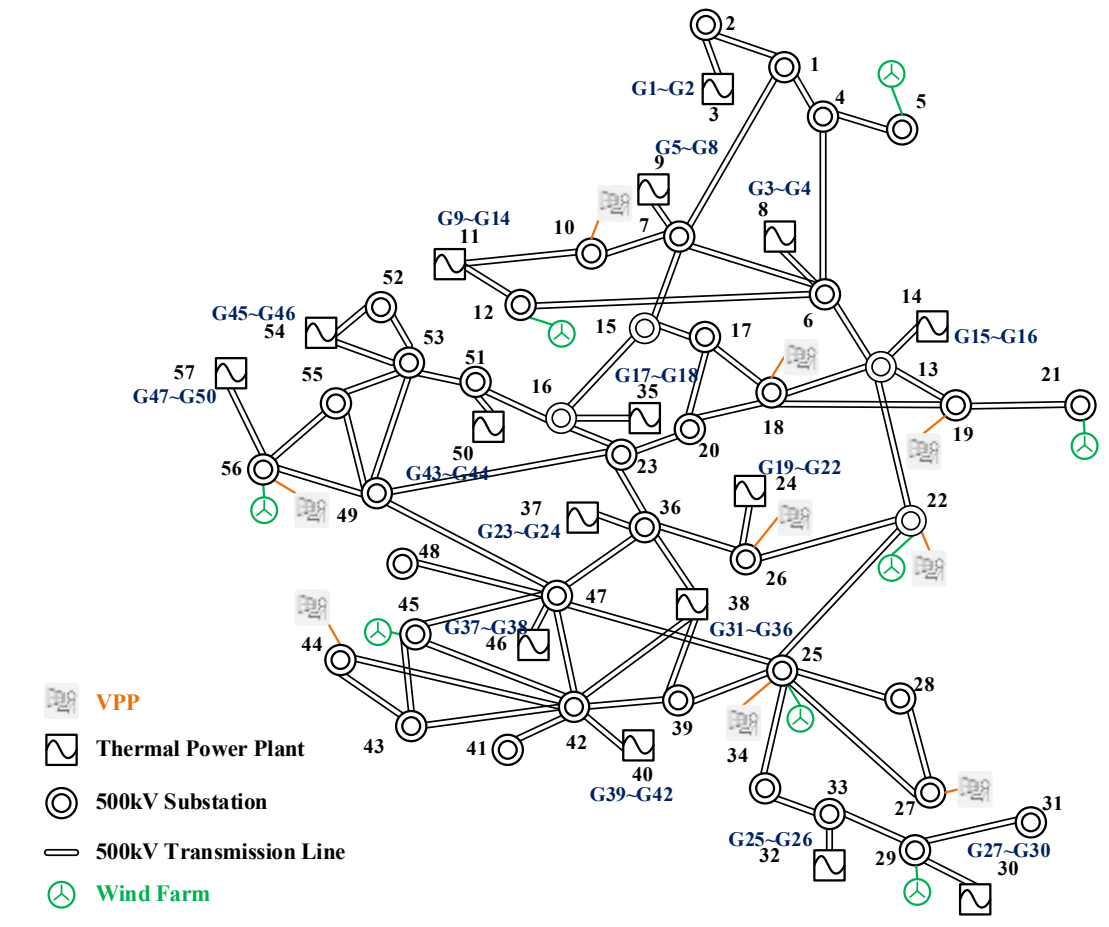

Fig. 12. Topological diagram of the real power system case.

## VIII. ACKNOWLEDGEMENT

We gratefully acknowledge Professor Cong Chen with the Dartmouth College for her invaluable insights and stimulating discussions. We also thank the anonymous reviewers for their constructive feedback which greatly enhanced this paper.

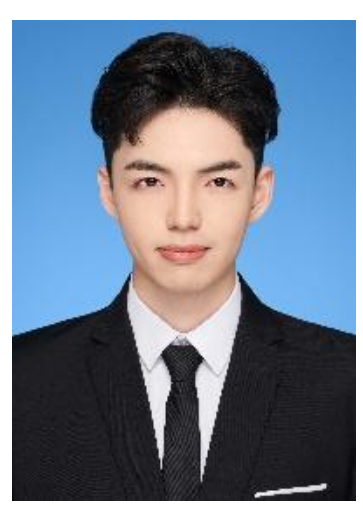

**Jingguan Liu** (Graduate Student Member, IEEE) received the bachelor's degree in electrical engineering from Huazhong University of Science and Technology, Wuhan, China, in 2022, where he is currently pursuing the Ph.D. degreee, advised by Prof. Xiaomeng Ai and Prof. Jiakun Fang. Since September 2025, he has been a visiting student at Dartmouth College, advised by Prof. Cong Chen. His research focuses on demand-side aggregation, energy storage operation, and optimization under uncertainty.

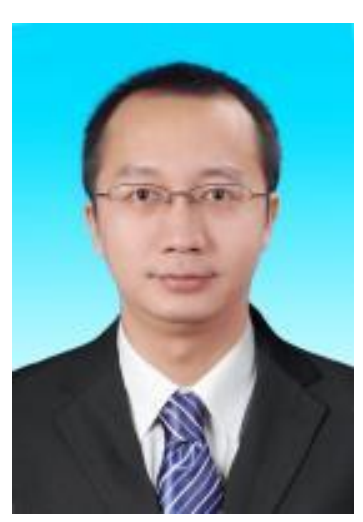

**Xiaomeng Ai** (Member, IEEE) received the B.Eng. degree in mathematics and applied mathematics and the Ph.D. degree in electrical engineering from the Huazhong University of Science and Technology (HUST), Wuhan, China, in 2008 and 2014 respectively. He is currently a Professor with the School of Electrical and Electronics Engineering, HUST. His research interests include robust optimization theory in power system, renewable energy integration, and integrated energy market.

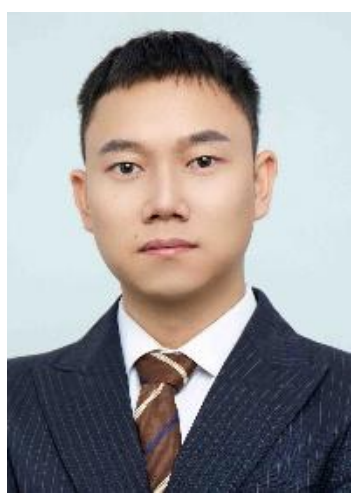

**Shichang Cui** (Member, IEEE) received the B.E. degree in automation and the Ph.D. degree in control science and engineering from the Huazhong University of Science and Technology (HUST), Wuhan, China, in 2016 and 2021, respectively. From 2019 to 2020, he was visiting the Department of Mechanical Engineering, University of Victoria, Canada, supported by the CSC Joint Doctoral Program. From 2021 to 2025, he was an assistant research fellow with the State Key Laboratory of Advanced Electromagnetic Technology (HUST), where he currently works as a lecture. His current research interests include energy management for smart grids, distributed optimization, and game theory.

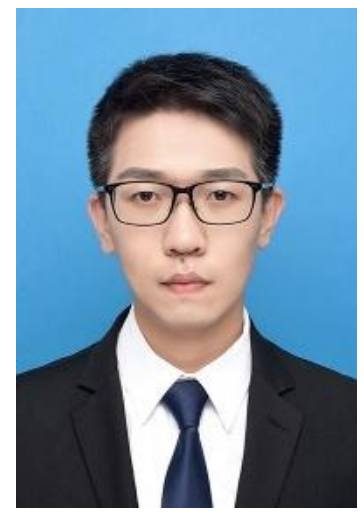

**Xizhen Xue** (Member, IEEE) received the B.E. and Ph.D. degrees in electrical engineering from the Huazhong University of Science and Technology, Wuhan, China, in 2019 and 2024, respectively. From 2023 to 2024, he was a Visiting Research Scholar with the Department of Electrical and Computer Engineering, Clarkson University, Potsdam, NY, USA. He is currently a Research Fellow with the School of Electrical and Electronic Engineering, Nanyang Technological University, Singapore. His current research interests include approximate dynamic programming, distributed optimization algorithm, energy storage scheduling and planning, and integrated energy systems.

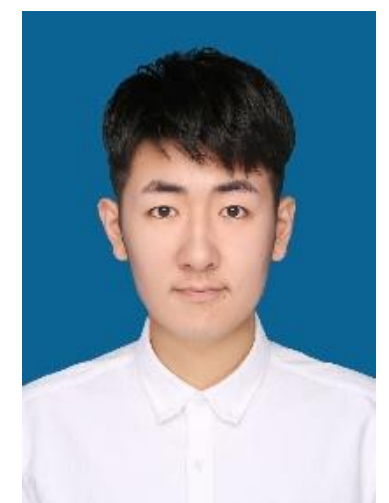

**Shengshi Wang** (Member, IEEE) received the B.Eng. degree in electrical engineering and automation from Chongqing University, Chongqing, China, in June 2020 and the Ph.D. degree in electrical engineering from Huazhong University of Science and Technology, Wuhan, China, in June 2025. From September 2023 to November 2024, he was a visiting student at Cardiff University, Cardiff, United Kingdom. Since September 2025, he has been a research fellow with the Singapore Institute of Technology, Singapore, which innovates closely with industry to pilot engineering solutions. His research interests lie in data-driven optimization—particularly robust optimization with decision-dependent uncertainties, approximate dynamic programming, and learning to optimize, with a focus on their applications in the energy systems.

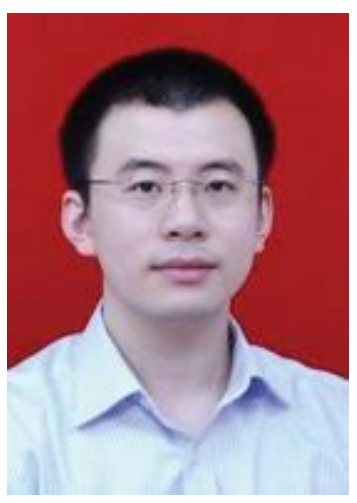

**Jiakun Fang** (Senior Member, IEEE) received the B.E. and Ph.D. degrees from the Huazhong University of Science and Technology (HUST), Wuhan, China, in 2007 and 2012, respectively. From 2012 to 2019, he was with the Department of Energy Technology, Aalborg University, Aalborg, Denmark. He is currently a Professor with the School of Electrical and Electronics Engineering, Huazhong University of Science and Technology. His research interests include the optimal integration of the power and gas systems, and the storage across multiple energy carriers.

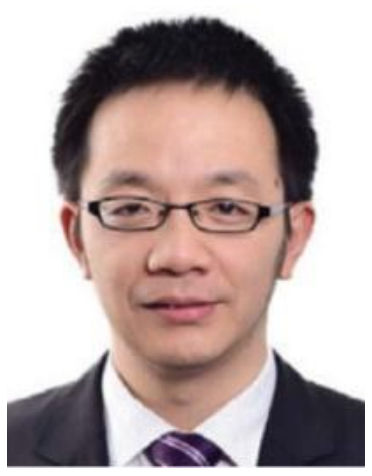

**Wei Yao** (Senior Member, IEEE) received the B.S. and Ph.D. degrees in electrical engineering from the Huazhong University of Science and Technology (HUST), Wuhan, China, in 2004 and 2010, respectively. He was a Postdoctoral Researcher with the Department of Power Engineering, HUST, from 2010 to 2012 and a Postdoctoral Research Associate with the Department of Electrical Engineering and Electronics, University of Liverpool, Liverpool, U.K., from 2012 to 2014. He is currently a Professor with the School of Electrical and Electronics Engineering, HUST. His current research interests include power system stability analysis and control, renewable energy, HVDC and DC Grid, and application of artificial intelligence in smart grid.

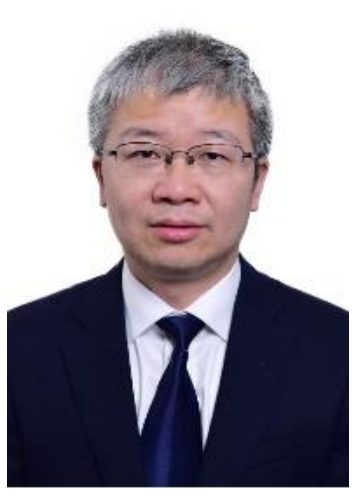

**Jinyu Wen** (Member, IEEE) received the B.S. and Ph.D. degrees in electrical engineering from the Huazhong University of Science and Technology, Wuhan, China, in 1992 and 1998, respectively. He was a Visiting Student from 1996 to 1997, and a Research Fellow from 2002 to 2003, with the University of Liverpool, Liverpool, U.K., and a Senior Visiting Researcher with the University of Texas at Arlington, Arlington, TX, USA, in 2010. From 1998 to 2002, he was a Director Engineer with XJ Electric Company Ltd., China. In 2003, he joined HUST, where he is currently a Professor with the School of Electrical and Electronics Engineering. His current research interests include renewable energy integration, energy storage, multiterminal HVDC, and power system operation and control.